\documentclass[prd,onecolumn,preprint,showpacs,12pt]{revtex4}
\usepackage{amsmath}
\begin{document}

\title{Almost Pure Phase Mass Matrices from Six Dimensions}
\author{P.Q. Hung}
\email[]{pqh@virginia.edu}
\author{M. Seco}
\email[]{Marcos.Seco@virginia.edu}
\affiliation{Dept. of Physics, University of Virginia, \\
382 McCormick Road, P. O. Box 400714, Charlottesville, Virginia 22904-4714}
%\date{\today}
\begin{abstract}
A model of quark masses and mixing angles is constructed within the framework
of two large extra compact dimensions. A ``democratic'' almost
pure phase mass matrix arises in
a rather interesting way. This type of mass matrix has often been used
as a phenomenologically viable ansatz, albeit one which had very little
dynamical justification. It turns out that the idea of Large Extra Dimensions
provides a fresh look at this interesting phenomenological ansatz as
presented in this paper. Some possible interesting connections to 
the strong CP problem will also be presented.
\end{abstract}
\maketitle
% body of paper here
The question of the origin of fermion mass hierarchy, mixing angles
and CP violating phase
is one of the most outstanding problems in particle physics. There have
been numerous attempts to study this problem, some of which are more
theoretical in nature while others are more phenomenological. However, it 
is generally agreed that the final word is far from being said. 
Furthermore, it is also agreed that the solution, whatever it might be,
is to be found outside of the Standard Model (SM).
 
In all of these studies, the phenomenological-ansatz approach is much
more modest in scope. Starting with some simple assumption about the
form of the mass matrix whose theoretical justification is
yet-to-be-determined, one could fit quark masses and mixing
angles. One of such approaches is particularly appealing: The pure
phase mass matrix (PPMM)~\cite{branco,Fishbane:xa}. This particular
ansatz is based on a simple assumption that there is a single and
unique Yukawa coupling for each quark sector and that the $3 \times 3$
mass matrix takes the form ${\cal M} = g_{Y}(v/\sqrt{2})\{
\exp(i\theta_{ij})\}$, where $i,j=1,2,3$. This kind of mass matrices
belongs to a class of the so-called ``democratic mass matrices''
(DMM)~\cite{dmm}. The pure phase mass matrix is attractive in that the
hierarchy of masses is governed by a single Yukawa coupling in the
limit where all phases vanish. A realistic hierarchy comes about when
the phases, which are treated as small perturbations, are put back
in. Although it is conceptually attractive, no attempt was made to
justify its underlying assumption.  Earlier works on trying to model
the pure phase mass matrix relied entirely on the framework of
four-dimensional field theories.  Although there are a number of
useful lessons that can be learned from this mode of thinking, one is
sometimes faced with more questions than answers.

On another front, there has been important conceptual developments in
the last few years related to a possible existence of Large Extra
Dimensions~\cite{Arkani-Hamed:1998rs,Antoniadis:1990ew}. 
Not only does this concept
force us to rethink about notions such as the question of what the
ultimate fundamental scale of nature might be, it also inspires us to
reformulate some of the longstanding problems in particle physics such
as the origin of fermion masses and mixings. The hierarchy of masses
has been reexamined recently within the framework of large extra
dimensions, and new interesting ideas have emerged such as the notion
of ``thick branes'' and the localization of various fermions inside
these branes~\cite{Arkani-Hamed:1999dc}. This localization can be
accomplished by a domain wall inside the brane.  This gave rise to the
idea of the strength of the Yukawa coupling (which is proportional to
the mass of the fermion) as being the overlap of the wave functions of
the localized fermions. As stated in Ref.~\cite{Arkani-Hamed:1999dc}, it is
easy to think of the reason why some fermions are heavy and some are
light: The heavy ones have large overlap and the light ones have small
overlaps.  There has been some works done along that line in order to
explain the fermion mass hierarchies. Most of these works made use of
the size of the wave function overlaps to discuss the fermion mass
problem.

Whatever various scenarios might be, the common important elements
which transpired from these works are basically the locations
of the domain walls and the size of the wave function overlaps. 
In fact, many of the physics results will depend on the actual
placements of the domain walls along the extra dimensions.

Our approach in this paper is as follows:
For each fermion sector (e.g. the up and down quark sectors),
there is a universal overall mass scale whose 
Yukawa coupling strength is determined by 
the size of the overlap. This
gives rise to a democratic mass matrix whose elements are
all equal to unity, apart from a common mass scale factor 
multiplied by an effective Yukawa coupling. 
All that is needed is to localize all the
left-handed fermions at one location, regardless of family indices,
and all the right-handed fermions at another location along the
fifth dimension inside the thick brane, and, in addition, 
to endow the fermions with a permutation symmetry.
Unfortunately, it is
well known that this kind of matrix does not work: one obtains
{\em one non-zero} mass eigenvalue and {\em two zero} eigenvalues.
The matrix $\{1\}$ has to be replaced by another quasi-democratic
one of the form such as $\{\exp(i\theta_{ij})\}$ for example.
The mass hierarchy which arises {\em within} each sector is
due, in our scenario, to the introduction of a sixth dimension and
a thick brane along it.
The introduction of ``family'' domain walls at different locations
inside this thick brane generate different phases for different
families. It will be 
seen that it is these phase differences which give rise to the
{\em pure phase} mass matrix or, as we shall see, an almost-pure
phase mass matrix.

We would like to make the following remark. Our model will contain
a certain number of parameters that need to be fixed phenomenologically.
However, what we present here is a new perspective on an old
problem which, hopefully, can give further insights 
which might be useful for future investigations. What we are
doing here is to try to
rephrase the origin of quark mass hierarchy (and eventually
that of the leptons as well) and CP phase in a completely new
context: that of the Compact Extra Dimensions (CED). We will show below
that the appearance of the phases in the mass matrices,
a crucial element in their construction, appear rather
``naturally''. From this point of view, it appears to be
a definite conceptual advantage of the CED scenario.

One remark is in order here concerning the introduction of a sixth
dimension. It is well known that, with just one extra compact
dimension, the fundamental 5-dimensional Planck scale cannot be
of the order of a few TeV or so, for it will introduce deviations
to the inverse square law on astronomical distances. Recent gravity
experiments~\cite{Hoyle:2000cv} down to a millimeter or so put a 
lower bound of
around 3 TeV on the 4+n Planck scale for the case of n=2
(with equal compactification radii). This fact, of course,
was not the one motivating us in introducing a sixth dimension. It 
is rather the natural way in which phase differences appear between
different fermions eventually giving rise to a pure phase mass matrix
which motivated us.

The organization of the paper is as follows. First we review
various features of fermions in five dimensions, including,
for instance, the concept of fermion localization. We then
show how, with a rather simple assumption, a democratic
mass matrix appears. Next, we introduce fermions in six 
dimensions and show how phase differences appear, and how
one can construct an (almost) pure phase mass matrix 
from this result. In this construction, ``family''
domain walls are introduced and it is shown that
their small separations along the sixth dimension are
responsible for the aforementioned phase differences.
Unlike what happens along the fifth dimension, the fermion
wave functions are not of the localizing type but are rather
oscillating. We will then discuss
how hermitian and non-hermitian pure phase mass matrices arise.
Finally, we will discuss some possible connections to the
strong CP problem~\cite{CP}.

\section{Fermions in 5 dimensions and Democratic Mass Matrix}

\subsection{A Review}

In this section, we will review some aspects of fermions in five
dimensions which have support $[0,L]$ along the fifth dimension.
In other words, we are discussing a ``thick brane'' of thickness
$L$. This discussion serves two purposes: to set the notations
and to lead to the democratic mass matrix.

We will adopt the effective field theory approach of
Refs. \cite{Georgi:2000wb,Cheng:1999bg}. This approach has the merit of being
relatively simple and transparent as far as the physics is concerned.
We first summarize below what has been done for the case of one flavor
of fermions, without and with a background scalar field.

To set the notations straight, the 4-dimensional coordinates will be
labeled by $x^{\mu}$ with $\mu = 0,..,3$ while the fifth coordinate
will be labeled by $y$.
We start out with a free Dirac spinor of $SO(4,1)$ which has four
components, $\psi$. The gamma matrices are $\gamma^{\mu}$ and
$\gamma_{y} = i \gamma_5$. The free Dirac Lagrangian is given by
\begin{eqnarray}
{\cal L}& = & \bar{\psi} \left(i\not\!\partial + i\gamma_{y}\frac{\partial}
{\partial y}\right)\psi, \\ \nonumber
        & = & \bar{\psi} \left(i\not\!\partial - \gamma_{5}\frac{\partial}
{\partial y}\right)\psi.  
\end{eqnarray}

The above Lagrangian has the following $Z_2$ symmetry:
$\psi (x, y) \rightarrow \Psi (x , y) = \pm \gamma_{5} \psi (x, L- y)$.
When this symmetry is combined with the periodic boundary
condition: $\psi (x, y) = \Psi (x, L+y) = \psi (x, 2 L + y)$,
one obtains: $\psi (x, -y) = \Psi (x,L- y)= \pm \gamma_{5}\psi (x, y)$
and $\psi (x, L + y) = \Psi (x, y)= \pm \gamma_{5}\psi (x, L- y)$,
which shows that $y=0,L$ are fixed points. One can subsequently
define the chiral components of $\psi$ by using the usual
operators $P_{R, L} = (1\pm \gamma_{5})/2$, with
$\psi_{+} = P_{R} \psi$ and $\psi_{-} = P_{L} \psi$, with
$\gamma_{5} \psi_{\pm} = \pm \psi_{\pm}$. The previous symmetry and 
boundary conditions are what usually referred to in the literature
as compactification on an $S_{1}/Z_{2}$ orbifold. One can
have fermions which have the symmetry
$\psi (x, y) \rightarrow \Psi (x , y) = +\gamma_{5} \psi (x, L- y)$,
and those which have
$\psi (x, y) \rightarrow \Psi (x , y) = -\gamma_{5} \psi (x, L- y)$.

For simplicity, we shall discuss the case
$\psi (x, y) \rightarrow \Psi (x , y) = +\gamma_{5} \psi (x, L- y)$
below. This corresponds to the case where only {\em right-handed}
zero modes survive in the brane, as shown below. For the other
situation,
$\psi (x, y) \rightarrow \Psi (x , y) = -\gamma_{5} \psi (x, L- y)$,
only the {\em left-handed} zero modes survive inside the brane,
as one can easily check.

Zero modes residing in the brane are supposed to be independent
of the extra coordinate, $y$ in this case. From the above discussion,
one can see that $\psi_{-}$ vanishes at the fixed points, and hence
there is no zero mode for $\psi_{-}$. The only non-vanishing zero mode
is $\psi_{0+}$. This can also be seen explicitly by writing
\begin{subequations} 
\begin{equation}
\label{poseig}
\psi_{M+} (x, y) = \psi_{M+} (x) \xi_{M+} (y),
\end{equation}
\begin{equation}
\label{negeig}
\psi_{M-} (x, y) = \psi_{M-} (x) \xi_{M-} (y),
\end{equation}
\end{subequations}
for a mode of mass $M$. From the explicit solutions for $\xi$ as given
in Ref. \cite{Georgi:2000wb}, one can again see that there is only one chiral
zero mode inside the brane. Four-dimensional chirality is seen to
arise from the symmetry and boundary conditions.  The chiral zero mode
$\psi_{0+}$ is {\em uniformly} spread over the fifth dimension $y$. To
localize $\psi_{0+}$ at specific points along $y$ inside the brane,
the use of domain walls have been suggested by
Refs. \cite{Georgi:2000wb,Arkani-Hamed:1999dc}. To this end, a background scalar
field, $\Phi$, is introduced. The Lagrangian is given by
\begin{eqnarray}
\label{lag}
{\cal L} & = & \bar{\psi} \left(i\not\!\partial - \gamma_{5}\frac{\partial}
{\partial y} - f \Phi\right)\psi + \frac{1}{2} \partial^{\mu} \Phi
\partial_{\mu} \Phi \\ \nonumber
         &  & -\frac{1}{2}\partial_{y}\Phi \partial_{y} \Phi
-\frac{\lambda}{4} (\Phi^2 - V^2)^2 .
\end{eqnarray}
The symmetry and boundary conditions on $\Phi$ are now:
$\Phi \rightarrow \tilde{\Phi}(x,L-y) = -\Phi(x,y)$;
$\Phi(x,-y) = \tilde{\Phi}(x, L-y) = -\Phi(x,y)$ and 
$\Phi(x,L+y) = \tilde{\Phi}(x, y) = -\Phi(x,L-y)$.
It can then be seen that $\phi$ vanishes at the orbifold fixed
points: $y=0,L$. As discussed in Ref. \cite{Georgi:2000wb}, $\Phi$
has a minimum energy configuration: $\langle\Phi (x,y)\rangle = \phi(y)$,
with $\phi(0) = \phi(L) = 0$. From the modified equations for
$\xi_{M\pm}$ with an added term $f \phi(y)$, one can easily
see the localization of the zero mode, namely 
\begin{equation}
\xi_{0+}(y) = k e^{-s(y)}, \xi_{0-}(y) = 0,
\end{equation}
where
\begin{equation}
s(y) = f \int_{0}^{y} dy^{\prime} \phi(y^{\prime}).
\end{equation}
As pointed out by Ref. \cite{Georgi:2000wb}, the chiral zero mode,
$\xi_{0+}(y)$, is now localized either at $y=0$ or $y=L$
depending on the sign of $f \phi(y)$. 

As in Ref. \cite{Arkani-Hamed:1999dc}, the special choice
$f \phi(y) = 2 \mu^2 y$ which makes the operators
$a= \partial_{y} + f \phi(y)$ and $a^{\dagger}= -\partial_{y} + f \phi(y)$
behave like the annihilation and creation operators of a Simple
Harmonic Oscillator (SHO), the normalized wave function for the
chiral zero mode $\xi_{0+}(y)$ takes on the familiar form
$\xi_{0+}(y) = (\sqrt{\mu}/(\pi/2)^{1/4})\exp( - \mu^2 y^2)$.
One clearly notices the localization of $\xi_{0+}(y)$ at $y=0$.
Another way of describing this phenomenon is the fact that
$\phi$ has a kink solution of the form $V \tanh((\lambda/2)^{1/2}
V y)$ which basically traps the fermion to a domain wall of size
$((\lambda/2)^{1/2} V)^{-1}$~\cite{Kaplan:1992bt}.

The next question concerns the possibility of localizing the chiral
zero mode at some other location than the one at the orbifold fixed
points. Ref. \cite{Arkani-Hamed:1999dc} has proposed to change the Yukawa
interaction $\bar{\psi} ( f\phi(y) ) \psi$ to $\bar{\psi} ( f\phi(y)-
m) \psi$ so that the wave function of the chiral fermion field is now
localized at the {\em zero} of $f\phi(y)- m$ instead of
$f\phi(y)$. With the SHO approximation, this zero would be at $y =
m/2\mu^2$. However, in order to be compatible with the $Z_2$ symmetry
of the Lagrangian, as shown in Eq. (\ref{lag}), one should also
require a ``mass reversal'' $m \rightarrow -m$ simultaneously with the
$Z_2$ transformations. This is the assumption we will be making in
this manuscript.  (Another approach is given in Ref. \cite{Georgi:2000wb}).

As emphasized by Ref. \cite{Arkani-Hamed:1999dc}, different massless
chiral fermions can be localized on different slices along $y$,
inside the thick brane. These locations are determined by the
zeros of $f\phi- m_{i} = 0$. Within the SHO approximation, the
wave functions are given by $(\sqrt{\mu}/(\pi/2)^{1/4})
\exp( - \mu^2 (y - y_{i})^2)$, where $y_i = m_{i}/2\mu^2$.
The interesting idea proposed in Ref. \cite{Arkani-Hamed:1999dc} is that
the effective Yukawa couplings between SM fermions and SM Higgs scalar,
which eventually determines the size of the mass term, are
mainly determined by the wave function overlap between the
left-handed and right-handed fermions. Hierarchy of masses
then appears to depend on the size of the overlaps. 

From hereon, we shall turn our attention to {\em left-handed}
zero modes inside the brane as used in the SM. As we have 
mentioned earlier, these come from five-dimensional fermions
with the $Z_2$ symmetry
$\psi (x, y) \rightarrow \Psi (x , y) = -\gamma_{5} \psi (x, L- y)$.

To prepare the groundwork for our subsequent discussion, let us 
write down the action in five dimensions of
a left-handed fermion, a right-handed fermion, and the Yukawa
interactions with a background scalar field, and a SM Higgs field.
Following Ref. \cite{Arkani-Hamed:1999dc}, we will denote
quarks in five dimensions by
the five-dimensional Dirac fields: $(Q, U^{c}, D^{c})$ and 
their {\em left-handed} zero modes by the following Weyl fields:
$(q, u^{c}, d^{c})$. 
Notice that with this notation,
a right-handed down quark, for example, will be $\bar{d^{c}}$.
Since we will be dealing in this
paper solely with the quark sector, we are not writing down the
lepton fields. This will be dealt with in a subsequent paper.
The SM transformations of the above fields are self-evident by
the use of these notations. In addition, one also introduces
two sets of scalar fields: a SM singlet background scalar field,
$\phi$, whose VEV is $\langle\Phi (x,y)\rangle = \phi(y)$, a SM doublet Higgs
field $H(x,y)$ whose zero mode $h(x)$ is assumed to be uniformly spread 
along $y$ inside the thick brane. The 5-dimensional action can be 
written as
\begin{eqnarray}
\label{action5}
S &=& \int d^{5}x\, \bar{Q} (i\not\!\partial_{5} + f \phi(y)) Q +
\bar{U^c} (i\not\!\partial_{5} + f \phi(y) - m_{U})U^{c} \\ \nonumber
&&+ \bar{D^c} (i\not\!\partial_{5} + f \phi(y) - m_{D})D^{c}+
\kappa_{U} Q^{T} C_5 H U^{c} + \kappa_{D} Q^{T} C_5 \tilde{H} D^{c},
\end{eqnarray}
where $C_{5} = \gamma_{0} \gamma_{2} \gamma_{y}$. 
From the above equation, one notices that $Q, U^{c}, D^{c}$ are localized
at $y_Q=0$, $y_{U}= m_{U}/2 \mu^2$, $y_{D} = m_{D}/2\mu^2$ respectively. 
In principle, $m_{U}$ and $\kappa_{U}$ can be different from $m_{D}$ and
$\kappa_{D}$ respectively. However, as we can see below, it is
sufficient to have $m_{U} \neq m_{D}$ in order for the resulting
masses of up and down quarks to be different, even if $\kappa_{U} =
\kappa_{D}$. Assuming that the zero mode of $H$ is uniformly
spread over $y$ inside the thick brane, the 4-dimensional
effective action for the Yukawa interaction for the up quark can be written as
\begin{equation}
S= \int d^{4}x\, \kappa_{U} q^{T}(x) h(x) u^{c} \int dy\, \xi_{q}(y)
\xi_{u^{c}}(y),
\end{equation}
and similarly for the down quark. From the form of the wave functions,
one obtains the 4-dimensional effective Yukawa couplings for up and
down quarks as follows
\begin{equation}
\label{Yu}
g_{Y,u} = \kappa_{U} \exp(-\mu^2 y_{U}^2/2),
\end{equation}
\begin{equation}
\label{Yd}
g_{Y,d} = \kappa_{D} \exp(-\mu^2 y_{D}^2/2),
\end{equation}

Two remarks can be made concerning Eqs. (\ref{Yu}) and (\ref{Yd}). First
of all, as emphasized by Ref. \cite{Arkani-Hamed:1999dc}, even if $\kappa$'s are
of order unity, the effective Yukawa couplings can be quite small if
$\mu y_{U,D} \gg 1$. Basically, the size of the effective coupling is
sensitive to the relative distance between left and right-handed
quarks as compared with the characteristic thickness of the domain
walls. The second remark concerns the Yukawa couplings in five
dimensions, $\kappa_{U,D}$. In this new framework of large extra
dimensions, one has to separate the mechanism which separates
$g_{Y,u}$ from $g_{Y,d}$, already at the level of the 5-dimensional
action from that which separates $g_{Y,u}$ from $g_{Y,d}$ at an
effective field theory level in four dimensions due to different
localization points along the extra dimension inside the thick
brane. It might happen that the 5-dimensional action has an up-down
symmetry in the Yukawa sector which is broken down inside the
brane. We shall return to this question at the end of the paper.

\subsection{Democratic Mass Matrix}
\label{DMM}

Let us, for now, concentrate on just one sector, e.g. the up sector. 
Let us assume that there are three families. The fermion fields in 
five dimensions that we will be dealing with in this section will be 
$Q$ and $U^c$. As we shall see below, in order to obtain the DMM scenario,
we will put all the $Q$'s at one location along $y$ inside the thick brane,
and all the $U^c$'s at another location. With this simple assumption
and the assumption that the SM Higgs zero mode is uniformly spread inside the
thick brane, one can naively obtain the democratic mass matrix mentioned
above. However, with the gauge field zero
modes also spreading uniformly inside the thick brane, 
this will give rise to unwanted flavor-changing neutral current 
(FCNC) operators. A symmetry has to be imposed in order to avoid these
FCNCs.

A simple symmetry that one can use is a permutation symmetry among the
three families, for both $Q$ and $U^c$. One can have: $S_{3}^{Q} \otimes
S_{3}^{U^c}$, with $Q \rightarrow S_{3}^{Q} Q$ and
$U^c \rightarrow S_{3}^{U^c} U^c$. The background scalar field described
earlier $\phi(y)$ is a singlet under the above permutation group. (In
this way, one will see that all $Q$'s are localized at one place and all
$U^c$'s are localized at another place.) One can now include gauge 
interactions in the kinetic terms of (\ref{action5}) by making the
replacement $\not\!\partial_5 \rightarrow \not\!D_5$, namely
\begin{eqnarray}
\label{kinetic}
S_0 &=& \int d^{5}x\, \bar{Q} (i\not\!D_{5} + f \phi(y)) Q +
\bar{U^c} (i\not\!D_{5} + f \phi(y) - m_{U})U^{c} \\ \nonumber
&&+ \bar{D^c} (i\not\!D_{5} + f \phi(y) - m_{D})D^{c}.
\end{eqnarray}
It is simple to see that $S_0$ is invariant under the above permutation
symmetry. Eq. (\ref{kinetic}) also implies that all $Q$'s are localized at 
one place and all $U^c$'s are localized at another place.

Next, we wish to introduce a Yukawa interaction between the SM Higgs
scalar and $Q$ and $U^c$. First, we notice that a term such as
\begin{equation}
\label{yukawa}
{\cal L}_{yukawa} = \kappa_{U} Q^{T} C_5 H U^{c} + h.c.  .
\end{equation}
breaks the permutation symmetry since $Q$ and $U^c$ transform under
different groups.
If they were to transform under the {\em same} permutation
group, Eq. (\ref{yukawa}) would be an invariant. However, it would give
a mass matrix of the form
\begin{equation}
{\cal M} = g_{Y,u} \frac{v}{\sqrt{2}}
\left(\begin{array}{ccc}
1&0&0 \\
0&1&0 \\
0&0&1
\end{array}
\right)\ \, .
\end{equation}
which is not of the DMM type. It turns out that with $S_{3}^{Q} \otimes
S_{3}^{U^c}$, one can construct an {\em invariant} for each permutation
group: $\sum_{i} Q_{i}$ for $S_{3}^{Q}$, and
$\sum_{j} U^{c}_{j}$ for $S_{3}^{U^c}$  where $i=1,2,3$ 
and $j=1,2,3$ are family indices. From this, one can construct
an invariant action for the Yukawa interaction
\begin{equation}
\label{yukawa2}
S_{yukawa}= \int d^{5}x\, \kappa_{U} \sum_{i} Q_{i}^{T} C_5
H \sum_{j} U^{c}_{j} + h.c.
\end{equation} 
The effective action in four dimensions can now be written as
\begin{equation}
\label{4Deff}
S_{eff,Yukawa} = \int d^{4}x\, \kappa_{U} \sum_{i,j}
q^{T,i}(x) h(x) u^{c,j} \int dy \xi_{q}^{i}(y)
\xi_{u^{c}}^{j}(y) + h.c.
\end{equation}
Since all the $q_{i}$'s are located at the same place inside
the brane, and similarly for all the $u^{c}_i$, the wave function
overlap $\int dy \,\xi_{q}^{i}(y) \xi_{u^{c}}^{j}(y)$ is universal and
{\em independent} of $i,j$. With this, one can now rewrite Eq. 
(\ref{4Deff}) as
\begin{equation}
\label{4Deff2}
S_{eff,Yukawa} = \int d^{4}x\, g_{Y,u}
q^{T}(x) \left(\begin{array}{ccc}
1&1&1 \\
1&1&1 \\
1&1&1
\end{array}
\right)\
h(x) u^{c} + h.c. \, ,
\end{equation}
where $g_{Y,u}$ is given by Eq. (\ref{Yu}), $q^{T}=
(q^{T}_1, q^{T}_2, q^{T}_3)$ and similarly for $u^{c}(x)$.
From Eq. (\ref{4Deff2}), one obtains the democratic mass matrix
\begin{equation}
\label{demo}
{\cal M} = g_{Y,u} \frac{v}{\sqrt{2}}
\left(\begin{array}{ccc}
1&1&1 \\
1&1&1 \\
1&1&1
\end{array}
\right)\ \, .
\end{equation}
An important remark is in order here. The universal strength in
Eq. (\ref{demo}) depends on, besides the SM quantity $v/\sqrt{2}
\sim 175$ GeV, $g_{Y,u}$ which is a product of two factors:
the five-dimensional Yukawa coupling, $\kappa$, and 
the overlap of left-handed and right-handed fermion wave functions.
In this scenario and its extension presented below, it is this product
that is important, and not simply the size of the overlap.

As we have mentioned above, the above matrix can be brought by
a similarity transformation to a form
\begin{eqnarray}
\label{SIM}
{\cal M}^{\prime} &= & S {\cal M} S^{-1} \\ \nonumber
&= &g_{Y,u} \frac{v}{\sqrt{2}}
\left(\begin{array}{ccc}
0&0&0 \\
0&0&0 \\
0&0&3
\end{array}
\right)\ \, .
\end{eqnarray}
As one can see above, one needs to move beyond the DMM scenario in order
to obtain a more ``realistic'' mass matrix. This is what we propose to
do in the next section.

One might wonder what the distinctive feature a fifth dimension
has to give us in regards with the above problem. Could one not
obtain a similar result staying in just four dimensions? In
principle, the answer is yes. However,
it appears more attractive to think that,
once $q_{i}$ are lumped together at one place and $u^{c}_{j}$
are lumped at another place, one would obtain the DMM naturally.
It is interesting to envision a scenario in which the Yukawa 
couplings are as universal as the gauge couplings themselves,
with the possibility that the effective Yukawa couplings 
can be different from one another due to the different
overlaps between left and right fermions. (Gauge interactions
are chirality conserving and, as a result, the effective
gauge coupling with the gauge boson zero mode is the same
as the original coupling.)

The above discussion carries over to the down sector in a similar
fashion. Obviously, although attractive, this kind of democratic mass matrix
does not give the correct mass spectrum. An extension of DMM was
discussed by Ref. \cite{branco}, in which, instead of having one's
as matrix elements, one has pure phase factors such as $\exp (i
\theta_{ij})$. (The diagonal elements can be all unity by a
suitable redefinition of the quark phases.) Explicitly, a pure
phase mass matrix looks like $
{\cal M} = g_{Y} (v/\sqrt{2}) (\exp(i\theta_{ij}))$.

To construct a model for PPMM- even for the special case such
as a symmetric matrix, one usually requires a rather complicated
Higgs structure \cite{Fishbane:xa}. That is if one stays in four dimensions.
One might wonder if extra dimensions might help in this regards.
We have seen above how an additional dimension could help 
conceptually in
obtaining a democratic mass matrix. The question we ask is
the following: Could pure phases such as $\exp (i
\theta_{ij})$ arise from extra dimensions and not from
some kind of complicated Higgs sector? In particular, if
we keep the Higgs sector to a minimum (one Higgs), this phase
cannot come from the Yukawa coupling nor from the VEV of the SM
Higgs. We have seen that,
in five dimensions, a chiral zero mode has, as a part of its wave function,
$\xi (y)$ which behaves, upon being trapped by a domain wall,
like $\exp (-\mu^2 y^2)$. As we shall see below, by adding
another compact dimension (the sixth one), the phases appear
as the overlaps between wave functions of fermions which are
``trapped'' at different locations along the 6th dimension.
What this really means will be explored in the next section.

\section{Fermions in 6 dimensions and Pure Phase Mass Matrix}

Notwithstanding the string theory argument, there might be another
simpler motivation for the need of more than one extra spatial
dimension: If the fundamental $4+n$ ``Planck'' scale were of O(TeV) to
``solve'' the hierarchy problem, and if the $n$ extra dimensions were
to be compactified with the same radius $R$ then $n \geq 2$ in order
for $R$ to be in the submillimeter region as required by the lack of
deviation from the ordinary inverse square law down to about 0.2
mm~\cite{Hoyle:2000cv}. In our case, the above need is dictated by the
desire to build a more ``realistic'' mass matrix: the so-called pure
phase mass matrix or its almost-pure-phase
counterpart. (In this construction, we are not concerned about
whether or not the ultimate theory contains more than six
dimensions.)
To this end, we first study the behaviour of
fermions in six dimensions, subject to similar boundary conditions as
in the 5-dimensional case.

\subsection{Fermions in six dimensions}

The task of this section is to study fermions in six dimensions,
with the ultimate aim of obtaining massless chiral fermions in
four dimensions.

In order to discuss fermions in six dimensions, we first turn our attention
to the representation of gamma matrices for these fermions. Before we
begin the discussion, a few remarks concerning spinors in $SO(N)$ are
necessary.

%As we have seen above, the 4-dimensional Dirac fermion is real. In general,
%for $SO(2n)$, spinor representations (with dimension $2^{n-1}$)
%are real if $n$ is even, and complex if $n$ is odd. We are particularly
%interested in the case with two extra spatial dimensions, and hence
%on the group $SO(6)$, or more precisely on $SO(5,1)$. $SO(6)$ has
%two irreducible 4-component right-handed and left-handed spinors, 
%$\psi_{+}$ and $\psi_{-}$, which are complex conjugates of one another. 

We shall be working with the group $SO(5,1)$ that, as we discuss in
Appendix \ref{appA}, has two irreducible spinor representations of
dimension 4. We shall put $\psi_{+}$ and $\psi_{-}$ into a reducible
``Dirac'' spinor $\psi = (\psi_{+}, \psi_{-})$. The chiral
representation of the gamma matrices for $SO(5,1)$ is shown in
Appendix \ref{appA}. The notation for the coordinates will be similar
to the five-dimensional case, with the sixth dimension denoted by $z$,
namely $x_{N} = ( x_0, x_1, x_2, x_3, y, z)$.  The free Lagrangian for
$\psi$ is now written as
\begin{equation}
\label{6Dlag}
{\cal L}_{\psi} = i\, \bar{\psi} \Gamma^{N} \partial_{N} \psi \, ,
\end{equation}
where $N=0,1,2,3,y,z$. The metric used in this paper is simply
(-+++++).
It is useful to see explicitly the 
Lagrangian written in terms of the components of $\psi$. For
this purpose, we give the explicit forms for $\Gamma_{y}$ and
$\Gamma_{z}$ as can be seen from Appendix \ref{appA},
\begin{equation}
\Gamma_{y} = \left(\begin{array}{cc}
0&-i\,\gamma_5 \\
i\,\gamma_5&0
\end{array}
\right)\
\end{equation}
\begin{equation}
\Gamma_{z} = \left(\begin{array}{cc}
0&I\!\!I \\
I\!\!I&0
\end{array}
\right)\
\end{equation}
where $\gamma_5$ is the usual matrix encountered in four dimensions
and $I\!\!I$ is a $4 \times 4$ unit matrix. In addition, we also need
$\bar{\psi} = \psi^{\dagger} \Gamma_0 = (\bar{\psi}_{-},
-\bar{\psi}_{+})$.
Eq. (\ref{6Dlag}) can now be rewritten as
\begin{equation}
{\cal L}_{\psi} = -i\, \bar{\psi_{+}} \gamma^{\mu} \partial_{\mu} 
\psi_{+} - i\, \bar{\psi_{-}} \gamma^{\mu} \partial_{\mu} 
\psi_{-} + \bar{\psi_{+}} \gamma^{5} \partial_{y} 
\psi_{+} + \bar{\psi_{-}} \gamma^{5} \partial_{y} 
\psi_{-} -i\, \bar{\psi_{+}} \partial_{z} \psi_{+}
+ i\, \bar{\psi_{-}} \partial_{z} \psi_{-} \, .
\end{equation}
As we explain in the Appendix, the 4-dimensional kinetic terms (the first
two terms of the above equation) will acquire a plus sign when 
$\gamma^{\mu}$ are replaced by $\tilde{\gamma}^{\mu}$ which are
appropriate for the metric $(-+++)$ which is a remnant of the original
metric $(-+++++)$. The reader is strongly recommended to consult
Appendix \ref{appA} concurrently with this section 
in order to avoid confusion.

As in the case of the fifth dimension,
we will assume that the sixth dimension is compactified on an
orbifold $S_{1}/Z_{2}$. 
$\psi$ is assumed to have support $[0,L_6]$ along the sixth dimension.
We first discuss this $Z_{2}$ symmetry for
free fermions.

From Eq. (\ref{6Dlag}), one can see that the Lagrangian has the
following $Z_2$ symmetry:
\begin{equation}
\label{sym6}
\psi(x^{\alpha},z) \rightarrow \Psi(x^{\alpha}, z) = \Gamma_{z} 
\psi(x^{\alpha}, L_{6}-z) \, .
\end{equation}
With $\Gamma_z$ given above,
this symmetry translates into
\begin{equation}
\setlength{\arraycolsep}{2pt}
\label{sym62}
\begin{array}{rcl}
\psi_{+}(x^{\alpha},z)& \rightarrow& \Psi_{+}(x^{\alpha},z)=
\psi_{-}(x^{\alpha}, L_{6}-z)\, ,\\ 
\psi_{-}(x^{\alpha},z)& \rightarrow& \Psi_{-}(x^{\alpha},z)=
\psi_{+}(x^{\alpha}, L_{6}-z) \, .
\end{array}
\end{equation}

As with the five-dimensional case, our boundary condition is
\begin{equation}
\label{bound}
\psi_{\pm}(x^{\alpha},z)=\Psi_{\pm}(x^{\alpha},L_{6}+z)=
\psi_{\pm}(x^{\alpha},2L_{6}+z) \, .
\end{equation}
Again, combining (\ref{sym62}) with (\ref{bound}), one obtains
\begin{equation}
\label{bound2}
\psi_{\pm}(x^{\alpha},-z)=
\psi_{\mp}(x^{\alpha}, z) \, ,
\end{equation}
\begin{equation}
\label{bound3}
\psi_{\pm}(x^{\alpha},L_{6}-z)=
\psi_{\mp}(x^{\alpha},L_{6}+ z) \, .
\end{equation}
We immediately recognizes $z=0, L_{6}$ to be the fixed points of the orbifold.
It is convenient to rewrite $\psi_{\pm}$ as
\begin{equation}
\label{chi}
\psi_{\pm} = \frac{1}{\sqrt{2}} (\chi \pm  \, \eta) \, .
\end{equation}
In terms of $\chi$ and $\eta$, the boundary conditions become
\begin{equation}
\chi (x^{\alpha}, -z) = \chi(x^{\alpha}, z) \, ,
\end{equation}
\begin{equation}
\eta(x^{\alpha}, -z) = - \eta(x^{\alpha}, z) \, .
\end{equation}
\begin{equation}
\chi (x^{\alpha},L_{6} -z) = \chi(x^{\alpha},L_{6}+ z) \, ,
\end{equation}
\begin{equation}
\eta(x^{\alpha},L_{6} -z) = -\eta(x^{\alpha},L_{6}+ z) \, .
\end{equation}
From the above boundary conditions, one can see that $\eta$ 
{\em vanishes} at the fixed points $z= 0, L_{6}$. 

As usual, we shall write:
\begin{subequations} 
\begin{equation}
\chi_{M} (x^{\alpha}, z) = \chi_{M} (x^{\alpha}) \xi_{\chi,M} (z),
\end{equation}
\begin{equation}
\eta_{M} (x^{\alpha}, z) = \eta_{M} (x^{\alpha}) \xi_{\eta,M} (z).
\end{equation}
\end{subequations}
Since the
{\em zero modes} in the ``4-brane'' are {\em independent} of $z$,
we have
\begin{equation}
\label{zero}
\chi(x^{\alpha}, z)_{0} = k \,\chi(x^{\alpha})\,;\,
\eta(x^{\alpha}, z)_{0} =0 \, ,
\end{equation}
where $k$ is a constant. Again, the free fermion wave function 
for the zero mode is
uniformly spread over the 6-th dimension. We now investigate 
the effect of
a coupling with a background scalar field having a kink solution.

For the discussion which follows, it is convenient to notice that 
\begin{equation}
\label{KE}
-i\, \bar{\psi_{+}} \partial_{z} \psi_{+}
+ i\, \bar{\psi_{-}} \partial_{z} \psi_{-} = 
i\,\bar{\chi} \partial_{z} \eta -i\, \bar{\eta} \partial_{z} \chi\,. 
\end{equation}
Eventually, we would like to find an equation for the surviving
zero mode $\chi(x^{\alpha}, z)_{0}$ in the presence of a
background scalar field which will be assumed to be real. For this
purpose, let us write the surviving zero mode $\chi$ as
\begin{equation}
\label{chi0}
\chi_{0}(x^{\alpha},z) = \chi(x^{\alpha})\xi_{\chi,0} (z)\,.
\end{equation}
As we shall see, upon using Eq. (\ref{KE}) and subsequent
interaction terms, one can derive an equation governing
the behaviour of $\xi_{\chi,0} (z)$ along $z$ which will
eventually tell us whether or not one has a localized behaviour
as in the five-dimensional case or an oscillatory one (pure
phase). This will depend on the type of fermion bilinears
which couple to the background scalar. Roughly speaking,
if the coupling ends up to be of the form 
$i\,\bar{\eta}\,\chi\,h(z)$, for example, then $\xi_{\chi,0} (z)$
will have an exponentially-suppressed form similar to the
five-dimensional case. If, however, it ends up looking like
$\bar{\eta}\,\chi\,h(z)$, then $\xi_{\chi,0} (z)$ will have
an oscillatory behaviour. This is so because of the way
Eq. (\ref{KE}) looks.

We now look for the aforementioned fermion bilinears which
are required to be hermitian (because the background scalar field
is assumed to be real) and Lorentz invariant.

%To study the above question, we shall keep in mind the following
%important feature of the class of models discussed in this
%paper and elsewhere: the boundary conditions as well as the
%existence of the kink solutions {\em break translational invariance}
%along the sixth dimension $z$. This allows us to build
%Lorentz-invariant interaction terms which might or might not
%be {\em local} along $z$. The nature of these interaction
%terms will depend on the transformation properties of the 
%background scalar field as we shall see below. 
%One also
%learns from a five-dimensional case (see e.g. Ref. (\cite{Georgi:2000wb})
%that, for finite $L_{6}$ as is the case here, the background
%scalar field can be written as a function of $z$ and $z^{\prime}
%= L_{6}-z$, namely $\phi = \phi(x^{\alpha},z, L_{6}-z)$.

Let us introduce a real scalar field
which transforms under $Z_{2}$ as
\begin{equation}
\label{Phi}
\Phi(x^{\alpha},z) \rightarrow  
-\Phi(x^{\alpha}, L_{6}-z) \, .
\end{equation}
%We now look for Lorentz-invariant and {\em hermitian} 
%bilinears which can couple to $\Phi(x^{\alpha},z)$. 
%We now
%present two ways of looking at this problem which give
%similar results. 
%We will make use of the fact that
%translational invariance is broken by the boundary conditions
%in this search.
%to list the following terms:

First, the most obvious, hermitian and Lorentz-invariant
bilinear is simply (remembering that $\Gamma_0$ is
anti-hermitian with our metric)
%\begin{subequations}
\begin{equation}
\label{comb1}
i\,\bar{\psi}(x^{\alpha},z)\psi(x^{\alpha},z) . 
\end{equation}
%\begin{equation}
%\label{comb2}
%i\,\bar{\psi}(x^{\alpha},z^{\prime})\psi(x^{\alpha},z^{\prime}) ; 
%\end{equation}
%\begin{equation}
%\label{comb3}
%\bar{\psi}(x^{\alpha},z)\psi(x^{\alpha},z^{\prime})-
%\bar{\psi}(x^{\alpha},z^{\prime})\psi(x^{\alpha},z) \, , 
%\end{equation}
%\end{subequations}
%where $z^{\prime} = L_{6} + z$. 
%The second combination
%(\ref{comb3}) is peculiar to a scenario with a compact dimension
%and vanishes in the limit $L_{6} \rightarrow \infty$.
Notice that (\ref{comb1}), when expanded in terms
of $\chi$ and $\eta$, are of the form $i\,\bar{\eta} \chi +...$.
This, when combined with Eq. 
(\ref{KE}), would give an exponentially-suppressed form for
the zero mode if there exists such a Yukawa coupling. Can it
couple to $\Phi$? If the reflection $Z_2$ symmetry were the
only symmetry around, it is straightforward to see that
a coupling of the form $i\,\bar{\psi}(x^{\alpha},z)\psi(x^{\alpha},z)
\Phi(x^{\alpha},z)$ is an invariant. This, as we have mentioned
above, would not be what we are looking for, namely an
oscillatory wave function. A mere mimicking of the five-dimensional
case would not work. Below we propose a mechanism where
the desired behaviour could arise.
%This is not the case for (\ref{comb3}).

Let us endow the scalar and fermion fields with an additional
discrete symmetry which will be called the $Q$-symmetry and 
which works as follows. Let us divide the space
inside the brane of thickness $L_6$ into two regions: $0$ to 
$L_{6}/2$ (Region I) and $L_{6}/2$ to $L_{6}$ (Region II).
Let us define the following transformations.
Under $Q$, 
\begin{equation}
\label{Qphi}
\Phi(x^{\alpha},z) \rightarrow -\Phi(x^{\alpha},z) \,.
\end{equation}
Notice that (\ref{Qphi}) is not to be confused with (\ref{Phi})
which is a reflection symmetry. We then notice the
following fact: If $z$ is inside Region I
then $L_{6}-z$ will be inside Region II and vice versa.
For the fermion, we will impose the following $Q$-transformations:
$\psi \rightarrow \psi$ for $z$ in Region I and
$\psi \rightarrow -\psi$ for $z$ in Region II.

With the above $Q$-symmetry, one notices that a coupling
of the form $\bar{\psi}(x^{\alpha},z)\psi(x^{\alpha},z)
\Phi(x^{\alpha}, z)$ is forbidden for any point $z$
inside the brane. However, a nonlocal interaction
of the form $\bar{\psi}(x^{\alpha},z)\psi(x^{\alpha},L_{6}-z)
\Phi(x^{\alpha}, z)$ is allowed by the $Q$-symmetry.
In particular, a {\em hermitian} bilinear containing
$\bar{\psi}(x^{\alpha},z)\psi(x^{\alpha},L_{6}-z)$ of
the form $\bar{\psi}(x^{\alpha},z)\psi(x^{\alpha},L_{6}-z)-
\bar{\psi}(x^{\alpha},L_{6}-z)\psi(x^{\alpha},z)$ is allowed by this symmetry.

The way the $Q$-symmetry works seems to imply that the orbifold
we used for the compactification should be $S_1/Z_2\times Z'_2$ 
instead of a $S_1/Z_2$. The behavior of the fields under the new
$Z'_2$ symmetry is, in fact, very similar to its behavior under 
the initial one. To see this let us define $z'=z-L_{6}/2$ and:
\begin{equation}
\widetilde\psi(x^\alpha,z')=\psi(x^\alpha,L_{6}/2+z')=\psi(x^\alpha,z)
\end{equation}
Again, from Eq. (\ref{6Dlag}), we can see that the Lagrangian is invariant
under the $Z'_2$ symmetry:
\begin{equation}
\label{z2psym}
\widetilde\psi(x^\alpha,z')\longrightarrow\widetilde\Psi(x^\alpha,z')=
\Gamma_z\widetilde\psi(x^\alpha,L_6-z')
\end{equation}
We will impose the same boundary condition as for $Z_2$:
\begin{equation}
\label{z2pbound}
\widetilde\psi(x^\alpha,z')=\widetilde\Psi(x^\alpha,L_6+z')
\end{equation}
Combining Eqs (\ref{z2psym}) and (\ref{z2pbound}) we get:
\begin{equation}
\setlength\arraycolsep{2pt}
\begin{array}{rcl}
\widetilde\psi_\pm(x^\alpha,-z')&=&\widetilde\psi_\mp(x^\alpha,z')\\
\widetilde\psi_\pm(x^\alpha,L_6-z')&=&\widetilde\psi_\mp(x^\alpha,L_6+z')
\end{array}
\end{equation}
which, in terms of $\psi$ and $z$, become:
\begin{equation}
\label{bound2b}
\begin{array}{rcl}
\psi_\pm(x^\alpha,z)&=&\psi_\mp(x^\alpha,L_6-z)\\
\psi_\pm(x^\alpha,-z)&=&\psi_\mp(x^\alpha,L_6+z)
\end{array}
\end{equation}

Using this second parity we can find an explicit realization of the 
$Q$-symmetry as follows. First, we shall define the behavior of the
fermions under this symmetry in the region I as,
\begin{equation}
\psi'(z)=Q\psi(z)=\Gamma_7\psi(z)
\end{equation}
now, Eq. (\ref{bound2b}) relates region I one and region II of the orbifold
-- as it should be since the physical space in a $S_1/Z_2\times Z'_2$ goes
from $0$ to $L/2$ -- so in order for $Q$ to be a symmetry of the Lagrangian
the fermions have to satisfy,
\begin{equation}
Q\psi(L_6-z)=\psi'(L_6-z)=\Gamma_z\psi'(z)=\Gamma_z\Gamma_7\psi(z)=
-\Gamma_7\psi(L_6-z),
\end{equation}
where in the second and last equalities we have used Eq. (\ref{bound2b})
which can also be written as $\psi(x^\alpha,z) = \Gamma_z \psi(x^\alpha,
L_{6}-z)$.

Notice that this realization of the $Q$-symmetry is only possible in
an even number of space-time dimensions since it is only in this case
that there exists a matrix which anticommutes with all of the gamma
matrices of the algebra and which does not belong to the algebra.

With the above definitions, it is straightforward to see
that the Yukawa coupling
\begin{equation}
\label{yuktilde}
{\cal L}_{Y} = f \,(\bar{\psi}(x^{\alpha},z)
\psi(x^{\alpha},L_{6}-z)-
\bar{\psi}(x^{\alpha},L_{6}-z)\psi(x^{\alpha},z))\,
\Phi(x^{\alpha},z) \, ,
\end{equation}
is invariant under all $Z_{2}$, $Z'_2$ and $Q$ symmetries
where $Q\,\Phi(x^{\alpha},z)=\Phi(x^{\alpha},z)$. 
Furthermore the action of the three parities forbids the 
presence of another non-local hermitian term,
$i\,(\bar\psi(z)\psi(L_6-z)+\bar\psi(L_6-z)\psi(z))\Phi(x^{\alpha},z)$.
In fact, Eq. \ref{bound2b} renders the above term to be
identical to zero.

In terms of $\chi$ and $\eta$. Eq. (\ref{yuktilde}) becomes
\begin{equation}
\label{lphix}
\setlength{\arraycolsep}{2pt}
\begin{array}{rcl}
{\cal L}_{Y1} &=& f\,\{(\bar{\chi}(x^{\alpha},z)
\eta(x^{\alpha},L_{6}-z) - \bar{\chi}(x^{\alpha},L_{6}-z)
\eta(x^{\alpha},z))\\
&+&(\bar{\eta}(x^{\alpha},z)\chi(x^{\alpha},L_{6}-z)-
\bar{\eta}(x^{\alpha},L_{6}-z)\chi(x^{\alpha},z))\}
\Phi(x^{\alpha},z)\,.
\end{array}
\end{equation}

As before, the minimum energy solution for $\Phi$ is
\begin{equation}
\langle\,\Phi\,\rangle = h(z) \, .
\end{equation}
From (\ref{KE}) and (\ref{lphix}), the equation of motion
for the surviving zero mode $\xi_{\chi,0} (z)$ has the form:
\begin{equation}
\label{zeromode}
-\partial_{z}\xi_{\chi,0} (z) + i\, f h(z) \xi_{\chi,0} (L_{6}-z) = 0 \, .
\end{equation}

In order to solve Eq. (\ref{zeromode}), we shall use Eq. (\ref{bound2b})
that, in terms of $\chi$ and $\eta$ leads to,
\begin{equation}
\label{ansatz}
\xi_{\chi,0} (L_{6}-z) = \xi_{\chi,0} (z) \, .
\end{equation}
%Notice that the above assumption is certainly true in the absence
%of the background scalar field (Eq. (\ref{zero})). We assume this 
%behaviour to carried over to the case {\em with} a background
%scalar field. (Another way of phrasing (\ref{ansatz}) is to say
%that we assume the reflection symmetry $z \leftrightarrow 
%L_{6}-z$ for the zero mode.)
Because
of the factor $i$ in Eq. (\ref{zeromode}) 
%and, as a result, we {\em do not expect} 
$\xi_{\chi,0} (z)$ {\em will not} be localized along $z$. 

%Before solving Eq. (\ref{zeromode}) using (\ref{ansatz}), let
%us describe another approach which yields results
%similar to the above scenario.

The solution to (\ref{zeromode}) with the ansatz
(\ref{ansatz}) is now given by
\begin{equation}
\label{chi1}
\xi_{\chi,0}(z) = \frac{1}{\sqrt{L}}e^{i s(z)}, 
\end{equation}
where
\begin{equation}
\label{phase}
s(z) = f \int_{0}^{z} dz^{\prime} h(z^{\prime}).
\end{equation}
Making the SHO approximation as used in the five dimensional case -a
statement to be justified below, the
properly normalized wave function for $\xi_{\chi,0}(z)$ would be
\begin{equation}
\label{chi2} 
\xi_{\chi,0}(z) = \frac{1}{\sqrt{L_{6}}}e^{i \mu^2 z^2} \, .
\end{equation}

From the above solution for the zero mode in the 6th dimension,
Eqs. (\ref{chi1}, \ref{chi2}) , we notice a marked difference
with the 5-dimensional case: the zero mode wave function is now
oscillating inside the thick brane, along the sixth dimension, while
in the five dimensional case, its counterpart has a localized form
along the fifth dimension. 

Let us assume there is a kink solution
for $\Phi$, i.e.
\begin{equation}
\label{kink}
h(z) = v \, \tanh (\mu z) \, ,
\end{equation}
where $\mu = (\lambda/2)^{1/2} v$. With this solution (\ref{kink}) put 
into (\ref{phase}), the
explicit expression for the non-vanishing zero mode is now
\begin{equation}
\label{chi4}
\xi_{\chi,0}(z) = \frac{1}{\sqrt{L_{6}}}e^{i f\,v\,\ln(
\cosh(\mu z))/\mu}, 
\end{equation}

Just as we have done with the five dimensional case, one could
generalize the above discussion to include a ``mass term'' so that $f
\, h(z) \rightarrow f \, h(z) -m$. As a result, one now has
\begin{equation}
\label{chi3}
\xi_{\chi,0}(z) = \frac{1}{\sqrt{L_{6}}}e^{i (f\,v\,\ln(
\cosh(\mu z))/\mu-m z)}\, . 
\end{equation}
This more general expression (\ref{chi3}) in fact determines
the phase of the oscillation.

%From the above discussions, one can visualize different fermions as
%having their phases ``localized'' along the sixth dimension (as
%determined by the ``zeros'' mentioned above). The precise meaning of
%``localization'' will be discussed in the next section. 
In the
construction of the mass matrices in four dimensions, we will need
overlaps of wave functions in the extra dimensions, as we have
discussed above in regards with the fifth dimension.  How the mass
matrices look like in six dimensions is the topic which will be
discussed next.

We end this section by presenting another type of Yukawa coupling
which is used to actually localize fermions along 
the fifth dimension. The
only difference with the previous section is that we now write it
using the full six dimensions. 
%We require that the background
%scalar field as well as the fermion bilinear to which it couples
%have even parities under the $Z_2$ symmetry of the sixth
%dimensions well as under the $Q$-symmetry.
With $\Gamma_7$
defined in Appendix A, the appropriate coupling is
\begin{equation}
S_{Yuk2} = \int d^{6}x f^{\prime} \bar{\psi} \Gamma_{7} \Phi^{\prime}
\psi \, .
\label{yuk2}
\end{equation}
Defining $\tilde{\gamma_{5}} = i\, \Gamma_{y} \Gamma_{7}$, one can
see that Eq. (\ref{yuk2}) is invariant under 
$\psi (x^{\mu}, y,z) \rightarrow \pm \tilde{\gamma_{5}} 
\psi(x^{\mu},L_{5} -y,
z)$ and $\Phi^{\prime}(x^{\mu},y,z) \rightarrow -\Phi^{\prime}(x^{\mu},
L_{5}-y,z)$ which finally gives
$\psi_{\pm} (x, -y,z) =  \pm \gamma_{5}\psi_{\mp} (x, y,z)$
and $\Phi^{\prime}(x,-y,z)= -\Phi^{\prime}(x,y,z)$. 
Also Eq. (\ref{yuk2}) is invariant under the $Q$ symmetry
provided that $Q\,\Phi^{\prime}(x^{\mu},y,z) = 
-\Phi^{\prime}(x^{\mu},y,z)$.
Notice that Eq. (\ref{yuk2}) can also be written as
\begin{equation}
S_{Yuk2} = \int d^{6}x f^{\prime} (\bar{\chi} \Phi^{\prime}
\chi - \bar{\eta} \Phi^{\prime} \eta)  \, .
\label{yuk3}
\end{equation}
Eq. (\ref{yuk3}) will reduce to the usual coupling in five dimensions.
One last commentis in order. Eq. (\ref{yuk2}) is also invariant under a
simultaneous $Z_2$-transformation:
$\psi(x^{\alpha},z) \rightarrow \Gamma_{z} 
\psi(x^{\alpha}, L_{6}-z) $,
$\Phi^{\prime}(x,y,z) \rightarrow \Phi^{\prime} (x,y,L_{6}-z)$,
as well as under the $Q$-symmetry.

Before leaving this section, we would like to make a remark concerning
Eq. (\ref{bound2b}). Basically, it is a ``mapping'' of region I into
region II and vice versa, namely $\psi(x^\alpha,z)= \Gamma_z \psi
(x^\alpha, L_{6}-z)$ or $\psi(x^\alpha,L_{6}-z)= \Gamma_z \psi
(x^\alpha, z)$. Now, let us remember that Eq. (\ref{bound2b}) is
a consequence of our boundary conditions. When we substitute it into
Eq. (\ref{yuktilde}) so that one deals with the
physical space which is now ranging from $0$ to
$L_{6}/2$, it acquires a Lorentz non-invariant form
$\bar{\psi}(x^\alpha,z)\Gamma_z \psi(x^\alpha,z)$. What this says is that
our boundary conditions break the six-dimensional Lorentz invariance
down to a five-dimensional Lorentz invariance. Our original
Lagrangian (\ref{yuktilde}) is {\em Lorentz invariant} under
the full six-dimensional Lorentz group and only when one goes to the 
physical space dictated by the boundary conditions, the
six-dimensional Lorentz invariance is broken down to the
five-dimensional one.

\subsection{(Almost) Pure Phase Mass Matrices}
\label{PPMM}

We shall use the same notations as in Section (\ref{DMM}). The
action for the Yukawa interaction, in six dimensions,
between the quarks and the SM Higgs field, is written
as (the Down sector is treated in exactly the same manner)
\begin{equation}
\label{yukawa6}
S_{yukawa}= \int d^{6}x \, \kappa_{U} \sum_{i} Q_{i}^{T} C_6
H \sum_{j} U^{c}_{j} + h.c. \, .
\end{equation} 
where $C_6=\Gamma_0\Gamma_2 \Gamma_z$.
We have, for the moment, omitted to write down other possible terms 
which are needed to determine the phases
along the sixth dimension. This will be dealt with in the next section.
We first begin with a ``phenomenological'' analysis.

The previous analysis led us to write a generic (zero-mode) fermion field as
\begin{equation}
\label{wfct}
\Psi (x,y,z) = \psi(x) \xi_{5}(y) \xi_{6}(z) \, ,
\end{equation}
Before making use of Eq. (\ref{yukawa6}) to construct the mass matrix,
let us describe a possible ``geography'' of the fermions along the
extra dimensions.  The discussion of Section (\ref{DMM}) pointed out
the following features: The localization, along the fifth dimension
$y$, of $Q_{i}$ at one place and $U^{c}_{i}$ at another place produces
a Democratic Mass Matrix as shown in Eq. (\ref{demo}). That is the
``geography'' along the fifth dimension that we would like to
keep. Basically, left-handed and right-handed fields are localized by
two domain walls at different locations. Why this should be so is
beyond the scope of this paper. However, one important point that
should be kept in mind is the fact that, in our model, there are only
two locations (left and right) along the fifth dimension, regardless
of the family index, for each quark sector (Up or Down). As mentioned
above, this gives rise to the universal effective Yukawa couplings
$g_{Y,u}$ and $g_{Y,d}$ which determine the overall mass strength for
each sector. Let us recall that $g_{Y,u}$ and $g_{Y,d}$ are
proportional to the overlap between left and right for the Up and Down
sectors respectively.  Again, what splits $g_{Y,u}$ from $g_{Y,d}$ is
beyond the scope of this paper. However, we will make some remarks
concerning this issue at the end of the paper.

The next question concerns the locations of various domain walls along
the sixth dimension. At the end of this section,
we will present a simple example which shows how one can localize
these domain walls. For the moment, we will simply parametrize these
locations as shown in Eq. (\ref{chi3}). We will assume that the domain
walls which ``fix'' the phases for the three families are located at 
different positions along $z$. For the purpose of illustration,
we will stay with this simple picture of family breaking in this
manuscript. A more general case with phenomenological applications
will be dealt with elsewhere. This will involve different profiles for 
different family kinks, etc..
%The important question is whether or
%not left and right (or $Q$ and $U^{c}$, e.g.) should be ``located''
%separately. We have seen that, with just one extra dimension ($y$),
%the effective Yukawa couplings depend on the the size of the 
%overlaps between $Q$ and $U^{c}$ wave functions along $y$. Should
%one expect a similar consideration to be made along the sixth
%dimension $z$ and what will the implications be?

We shall discuss below the implications of the cases when, for
each family, $Q$ and $U^{c}$ are ``in phase'' and when they are
slightly ``out of phase''. But, first, let us use Eq. (\ref{wfct})
and Eq. (\ref{yukawa6})
to construct a general generic mass matrix for the Up sector. The
mass matrix for the Down sector will be obtained in exactly the
same manner.

In the following, the quantity $L_{6}$ which appears in various
formulas is a generic symbol for the length of the physical
space, which is $L_6$ itself for the orbifold $S_1 /Z_2$ or
$L_{6}/2$ for the orbifold $S_1 /(Z_2 \times Z_2^{\prime})$.

To begin, we will assume the following situation for the
``geography'' of family domain walls along the sixth dimension $z$.
We will then discuss special cases of such a scenario. (As
we have briefly mentioned above, this scenario is presented for
the purpose of illustration and is not the most general case.)
Let us define the following
quantities which appear in Eq. (\ref{chi3}):
\begin{equation}
\label{def}
f \, v_{i}/\mu_{i} \equiv a_{i}\, ; \, m_{i;Q,U^{c}} 
\equiv m_{i,\mp} \, ,
\end{equation}
where $i=1,2,3$ denotes the family index and where $\mu_{i}
= (\lambda/2)^{1/2} v_{i}$. Notice that, in principle, the
quartic coupling $\lambda$ can depend on the family index
$i$. This more general case, however, will be investigated
elsewhere.
From Eqs. (\ref{yukawa6},~\ref{wfct}), one can write an effective
Yukawa interaction in four dimensions and construct a mass matrix
as we had done earlier. This construction is identical to the
five-dimensional case, except that now the matrix elements
will contain an extra factor which is the overlaps of $\xi_{6}(z)$'s.
As usual, the mass matrix will be similar to Eq. (\ref{demo}) except 
that now, instead of the matrix elements being unity, one has
\begin{equation}
\label{PPMM1}
{\cal M} = g_{Y,u} \frac{v}{\sqrt{2}}
\left(\begin{array}{ccc}
a_{11}&a_{12}&a_{13} \\
a_{21}&a_{22}&a_{23} \\
a_{31}&a_{32}&a_{33}
\end{array}
\right)\ \, ,
\end{equation}
where
\begin{subequations}
\begin{eqnarray}
\label{diag} 
a_{jj} & = & \int dz \, \xi_{6,j+}^{\ast} \xi_{6,j-} \\ \nonumber
& = &
\frac{1}{L_6} \int_{0}^{L_6} dz \, \exp(i (m_{j+} - m_{j-})z) \\ \nonumber
       & = & (exp(i (m_{j+} - m_{j-})L_{6})-1)/i (m_{j+} - m_{j-})L_{6} \, ,
\end{eqnarray}
\begin{eqnarray}
\label{offdiag} 
a_{ij} & = & \int dz \, \xi_{6,i+}^{\ast} \xi_{6,j-} \\ \nonumber 
& = &
\frac{1}{L_6} \int_{0}^{L_6} dz \, \exp(i(a_{j} \ln(\cosh(\mu_{j}z))-
a_{i} \ln(\cosh(\mu_{i}z))+ (m_{i+} - m_{j-})z)  \, .
\end{eqnarray}
\end{subequations}
Notice that $L_6$ here is a generic symbol for the length
of the physical space as we have mentioned above.

The above equations (\ref{PPMM1}, \ref{diag}, \ref{offdiag}) refer to
the case where domain walls, which ``determine'' the
phases of the fermions, are ``located'' at different places. We will 
specialize below to a few interesting possibilities. However, some
important remarks can already be made. We ask the following
question: Under what conditions will the mass matrix be
hermitian or non-hermitian?

\subsubsection{Hermitian and Non-Hermitian mass matrices}
\label{HMM}

We now present two different scenarios.

(a) The parameters $m_{i\pm}$ which determine the ``locations'' of
the domain walls possess interesting features. The first observation
one can make is as follows. If the domain walls which ``localize''
the phases of $Q$ and $U^{c}$ (Left and Right), for {\em each} family, 
are located at the same place along $z$, i.e.
\begin{equation}
\label{collapse}
m_{i+} = m_{i-} \,
\end{equation}
one obtains the following results
\begin{subequations}
\begin{equation}
a_{jj} = 1 \, ,
\end{equation}
\begin{equation}
a_{ji} = a_{ij}^{\ast} \, .
\end{equation}
\end{subequations}

The mass matrix ${\cal M}$ is {\em hermitian}! The hermiticity of the
mass matrix is a {\em consequence} of the ``collapse'' of left and
right (or $Q$ and $U^{c}$), for each family, into the ``same
position'' along the sixth dimension.  Two remarks can be made
concerning a hermitian matrix. First, its determinant is {\em
real}. This means that $\arg(\det {\cal M}) =0$.  The 
possible connection of this
statement with the strong CP problem (see e.g. a review by
~\cite{CP}) will be explored further at the
end of the paper. 

%Second, it has been shown by a number of
%phenomenological analyses~\cite{branco} that hermitian mass matrices
%produce unrealistic mass spectrum. In consequence, a deviation from
%hermiticity is necessary in order to produce a realistic mass
%spectrum. Within our framework, this means that one should
%``separate'' left and right (or $Q$ and $U^{c}$), i.e.  one should
%have $m_{i+} \neq m_{i-}$. On the other hand, the equality $m_{i+} =
%m_{i-}$ implies $\arg(\det {\cal M}) =0$. Could this be an attractive
%``solution'' to the strong CP problem? We shall take the point of view
%that the setting $m_{i+} = m_{i-}$ which gives rise to a hermitian
%matrix which, in turns, implies $\arg(\det {\cal M}) =0$, is a
%``tree-level'' condition. This is similar in spirit to other proposed
%alternative solutions to the strong CP problem, without making use of
%the axion, as reviewed in ~\cite{CP}.  We will return to these 
%issues below. 
Let us first see
if the hermitian matrix above is of a pure phase form.

The discussion
which follows will deal with issues which are also relevant to
the non-hermitian case.

Let us look at
\begin{equation}
\label{offdiag2} 
a_{ij}  = 
\frac{1}{L_6} \int_{0}^{L_6} dz \, \exp(i(a_{j} \ln(\cosh(\mu_{j}z))-
a_{i} \ln(\cosh(\mu_{i}z)) + (m_{i} - m_{j})z))  \, .
\end{equation}
Under what conditions would $a_{ij}$'s look like pure phases,
namely of the form $e^{i\theta}$, or an almost pure phase of
the form $(1-\rho)e^{i\theta}$ with $\rho \ll 1$? 
To answer this question,
let us make a little detour to the meaning of wave function
overlaps, thickness of domain walls and size of the extra dimensions.

We have seen how one can localize fermions along the fifth dimension
($y$) by having domain walls of sizes $1/\mu \ll L_5$. The
effective strengths of various interactions are determined
by the overlaps of the wave functions along $y$. For this reason,
it is preferable to have the thickness of the domain walls
small enough, i.e. $1/\mu \ll L_5$, so one can ``fit'' several fermions
along $y$ in such a way as to obtain desirable effects such as
``slow'' (or no) proton decay, possible mass hierarchies between
different fermion sectors (quarks, leptons), etc... As we move
on to the sixth dimension, it is not obvious that such a picture
is still necessary. In fact, at least as far as the pure phase
mass matrix is concerned, the thickness of these domain walls can
be as large as the size of the compactified dimension itself, as
we shall see below. 

Let us, for the time being, assume that all domain wall thicknesses
(along $z$) are of the size of the compact dimension, i.e.
$1/\mu_{i} \sim O(L_6)$. In this
situation, one can use the SHO approximation and carry out the
integration of Eq. (\ref{offdiag2}), namely
\begin{equation}
\label{offdiag3} 
a_{ij}  = 
\frac{1}{L_6} \int_{0}^{L_6} dz \, \exp(-i(\Delta \mu_{ij}^2 z^2 - 
\Delta m_{ij} z))  \, ,
\end{equation}
where
\begin{subequations}
\begin{equation}
\Delta \mu_{ij}^2 \equiv (1/2)(a_{i}\mu_{i}^2 -a_{j} \mu_{j}^2) \, ,
\end{equation}
\begin{equation}
\Delta m_{ij} \equiv m_{i} - m_{j} \, .
\end{equation}
\end{subequations}
The integration can be explicitly carried out. One obtains
\begin{equation}
\label{elij} 
a_{ij} = \frac{\sqrt{\pi}}{2} \frac{\text{erf}\left(\displaystyle
\frac{i(2\Delta \mu_{ij}^2 L_6
-\Delta m_{ij})}{2\sqrt{i\, \Delta \mu_{ij}^2}}\right) + 
\text{erf}\left(\displaystyle\frac{i\,\Delta m_{ij}}
{2 \sqrt{i\,\Delta \mu_{ij}^2}}\right)}
{\sqrt{i\, \Delta \mu_{ij}^2}\, L_6} \exp\left(i\, \frac{(\Delta m_{ij})^2}
{4\,\Delta \mu_{ij}^2}\right) \, .
\end{equation}
In a phenomenological application of Eq. (\ref{elij}), one can
use it without making any approximation. However, in order to
see if it has a more familiar pure phase form or not, we will
make an expansion of (\ref{elij}).

Let us define
\begin{equation}
\label{xij}
\sqrt{\Delta \mu_{ij}^2}\,L_6 \equiv x_{ij} \, ,
\end{equation}
\begin{equation}
\label{yij}
\Delta m_{ij} \, L_6 \equiv y_{ij} \,.
\end{equation}
For $x_{ij},\, y_{ij} < 1$, one can expand (\ref{elij}) giving
\begin{equation}
\label{elijnew}
a_{ij} = \{1-\frac{2}{45}x_{ij}^4 -\frac{1}{24}y_{ij}^2 + 
\frac{1}{12}x_{ij}^{2} y_{ij}\}\exp\{i(\frac{y_{ij}}{2}-
\frac{x_{ij}^2}{3})\} \, ,
\end{equation}
where we have neglected terms of $O(x_{ij}^8, y^4)$ or less in
the modulus and terms of $O(x_{ij}^6, y^4)$ in the phase.
Notice that for $a_{ji}$, one has $x_{ji}^2 = -\,x_{ij}^2$
and $y_{ji} = -\,y_{ij}$, and hence $a_{ji} = a_{ij}^{\ast}$
as they should. In this form one can see that 
the hermitian mass matrix is {\em almost} of the pure phase
form. This would have been the case if one could neglect
terms containing $x_{ij}$ and $y_{ij}$ inside the coefficient
multiplying the exponential. However, we will not neglect
those terms, leaving the possibility of a small deviation
\cite{Teshima:1996xc} from a pure phase mass matrix.

%Since we have assumed above that $L_{6} \sim 1/\mu_{i}$, we will
%concentrate on the following case:
%\begin{equation}
%\label{cond}
%\sqrt{\Delta \mu_{ij}^2} \, L_{6} \ll 1;\, \Delta m_{ij} \ll
%\sqrt{\Delta \mu_{ij}^2} \, .
%\end{equation}
%The second inequality in (\ref{cond}) expresses the assumption
%that the domain walls are separated from each other by distances
%which are much smaller than their thicknesses.
%Making use of the expansion $\text{erf}(x) = (2/\sqrt{\pi})x +O(x^3)$,
%one obtains
%\begin{equation}
%\label{elij2}
%a_{ij} \approx \exp\left(i\, \frac{(\Delta m_{ij})^2}
%{4\,\Delta \mu_{ij}^2}\right) \, .
%\end{equation}
%From Eq. (\ref{elij2}), one can see that the mass matrix is of a
%pure phase form. In addition, since $\Delta \mu_{ij}^2 = 
%-\Delta \mu_{ji}^2$, it is also hermitian. 

Notice that when
the domain walls are all located at the same point, i.e.
$\Delta m_{ij} = 0$, and when they have the
same thickness (or $\mu_{i} = \mu_{j}$),
$\forall i,j$, one recovers the DMM form,
namely $a_{ij} =1$, as one can see from Eq. 
(\ref{offdiag2}, \ref{offdiag3},\ref{elijnew}).
In addition, we notice that one can also obtain the
almost-pure phase hermitian mass matrix when either 
$\Delta m_{ij} \neq 0$ or $\Delta \mu_{ij}^2 \neq 0$,
but not necessarily both, as can easily be seen.

(b) As we have seen above, within
the framework of Eqs. (\ref{diag},\ref{offdiag}), the
mass matrix can be purely hermitian provided the
condition $m_{i+} = m_{i-}$ is fulfilled. What would
happen if $m_{i+} \neq m_{i-}$? To study this question,
let us refer back to Eqs. (\ref{diag},\ref{offdiag}, \ref{elijnew})
and let
\begin{equation}
\label{epsilon}
m_{i+} - m_{i-}= \epsilon_{i} \, .
\end{equation}
Also for convenience, let us define
\begin{equation}
\label{delta}
\delta_{i} = \epsilon_{i}\, L_{6} \, .
\end{equation}
With the above definitions, the diagonal matrix elements which
are no longer unity, can be written as
\begin{equation}
\label{aiinew}
a_{ii} = \exp(i\delta_{i}/2)\, \frac{\sin(\delta_{i}/2)}
{(\delta_{i}/2)} \, .
\end{equation}
The off-diagonal elements are similar to Eq. (\ref{elijnew}), except
that now one has the following replacement
$y_{ij} \rightarrow y_{i+,j-} = (m_{i+} - m_{j-})L_{6}$.
It is convenient to remove the phases from the diagonal elements by
absorbing the phases into $\xi_{6,i+}$, namely
$\xi_{6,i+} = \exp(i\delta_{i}/2)\,\xi_{6,i+}^{\prime}$. From
the definitions of $a_{ii}$ and $a_{ij}$, one now has
\begin{equation}
\label{aiinew2}
a_{ii} = \frac{\sin(\delta_{i}/2)}
{(\delta_{i}/2)} \, ,
\end{equation}
\begin{equation}
\label{elijnew2}
a_{ij} = \{1-\frac{2}{45}x_{ij}^4 -\frac{1}{24}y_{i-,j-}^2 + 
\frac{1}{12}x_{ij}^{2} y_{i-,j-}
-\frac{1}{12}y_{i-,j-}\delta_{i}-
\frac{1}{24}\delta_{i}^2 + \frac{1}{12}x_{ij}^{2}\delta_{i}\}
\exp\{i(\frac{y_{i-,j-}}{2}-
\frac{x_{ij}^2}{3})\} \, ,
\end{equation}
where 
\begin{equation}
\label{yij-}
y_{i-,j-} = (m_{i-} - m_{j-})L_{6} \, .
\end{equation}
Notice that $y_{i-,j-} = -y_{j-,i-}$.
In Eq. (\ref{elijnew2}), we have made use of the above phase
redefinition and of (\ref{epsilon},\ref{delta}). The mass matrix
described by the above elements is {\em not} hermitian for
the following reason. The modulus of $a_{ji}$ will have
a term $-\frac{1}{12}y_{j-,i-}\delta_{j}-
\frac{1}{24}\delta_{j}^2 + \frac{1}{12}x_{ji}^{2}\delta_{j}=
\frac{1}{12}y_{i-,j-}\delta_{j}-
\frac{1}{24}\delta_{j}^2 - \frac{1}{12}x_{ij}^{2}\delta_{j}$.
It can easily be seen that  $|a_{ij}| \neq |a_{ji}|$ unless
$\delta_{j} = -\delta_{i}$ which cannot be satisfied for
all $j$. Despite the fact that the phase of $a_{ji}$ is
the negative of that of $a_{ij}$, the difference in modulii
implies that, in general, $a_{ji} \neq a_{ij}^{\ast}$, and
hence the {\em non-hermiticity} of the matrix. It can be
approximately hermitian if one can neglect the terms containing
$\delta_{i}$ in (\ref{elijnew2}).

Notice that, even for the special case where all
``left-handed'' family domain walls are ``located'' at one point
along $z$, i.e. $m_{i-} = m_{j-} = m_{-}$, so that
$y_{i-,j-}=0$, and all
``right-handed'' family domain walls at another place, i.e.
$\delta_{i} = \delta$, the non-hermiticity still appears
in the difference in mudulii between  $a_{ji}$  and $a_{ij}$
because of the presence of $\delta$.

From the above discussion, one can see that one 
recovers the hermitian matrix in the limit
$\delta_{i} \rightarrow 0$.

In summary, we have shown that, in general, the deviation from
hermiticity in our framework comes from the splitting
between ``left'' and ``right'', namely $m_{i+} \neq m_{i-}$.

%(THIS IS THE TEMPORARY END OF THE MODIFICATION 4/25)
%(b) The second possibility where a hermitian mass matrix might arise
%is as follows. 
%One can put all the $Q$'s at  one location,
%namely $m_{i-} = m_{-}$, and all the $U^{c}$'s at another
%location, namely $m_{i+} = m_{+}$. Let us define
%\begin{equation}
%\label{del+-}
%(m_{i+} - m_{j-})^2 = (m_{+} - m_{-})^2 \equiv \Delta_{+-}^2 \,.
%\end{equation}
%With this arrangement, one now has
%\begin{subequations}
%\begin{equation}
%a_{jj} \approx 1 \, ,
%\end{equation}
%\begin{equation}
%\label{elij3}
%a_{ij} \approx \exp\left(i \frac{\Delta_{+-}^2}
%{4\,\Delta \mu_{ij}^2}\right) \, ,
%\end{equation}
%\end{subequations}
%where one now uses
%\begin{equation}
%\label{cond2}
%\sqrt{\Delta \mu_{ij}^2} L_{6} \ll 1;\, |\Delta_{+-}| \ll
%\sqrt{\Delta \mu_{ij}^2} \, .
%\end{equation}
%Again, one notices that the mass matrix is hermitian because
%of the fact that $\Delta \mu_{ij}^2 = -\Delta \mu_{ji}^2$.

The above analysis can be carried over to the Down sector in
exactly the same manner. There are however two interesting
remarks that can be made. First, although the mass matrix
for the Down sector is now characterized by a universal
strength $g_{Y,d}$ which is in general different from
$g_{Y,u}$, the matrix itself can be identical to the one
for the Up sector if we consider scenario (a). The reason
is that scenario (a) is one in which the domain walls
for $Q$ and $D^c$, for
each family, are ``located'' at the same place along the
sixth dimension, which is exactly the same as for the Up
sector. Therefore the matrix elements (without the
universal strength) are {\em the same}. In consequence,
the diagonalization matrices are {\em the same}, i.e.
$V_{U} \equiv V_{D}$. Hence, $V_{CKM} = V_{U}^{\dagger}
V_{D}= 1$, a mere unit matrix. In other words, the
mass matrices for the Up and Down sectors cannot be {\em both}
hermitian. To obtain a non-trivial CKM matrix, at least
one of the two matrices has to be non-hermitian in this
particular scenario.

The above (almost) pure phase mass matrix as obtained from six dimensions
is what we have set out to derive. From it, we have learned a few
things. 

(a) In general, the almost pure phase form of the mass matrix 
can be easily seen if the thickness
of various domain walls along the sixth dimension is of the order
of the compactified sixth dimension. (There is no reason why, in
principle, the thickness of the domain walls should be much smaller
than the compactified dimension, in contrast with the five-dimensional 
case.)

(b) When the domain walls ``fixing'' the phases for $Q$ and $U^{c}$,
{\em for each family}, are located at the same place,
($m_{u,i+} = m_{u,i-}$), the mass matrix is purely hermitian. 
As we have seen above, another possibility is when the domain
walls ``fixing'' the phases for $Q$ are at one location and
those which are responsible for ``fixing'' the phases of
$U^{c}$ are at another location, in which case the mass
matrix is also hermitian. If
one considers these cases to be a ``tree-level'' situation -a statement
to be further clarified below, the fact that 
$\arg(\det {\cal M}) =0$ makes this scenario an interesting
``candidate'' for a solution to the strong CP problem.

(c) The mass matrix becomes {\em non-hermitian} when 
$m_{u,i+} \neq m_{u,i-}$. 
%This non-hermitian pure phase mass matrix 
%is the desirable one, from the point of view of phenomenology.
We will briefly discuss below the possibility that
$m_{u,i+} \neq m_{u,i-}$ is due to ``radiative corrections'' of the
case $m_{u,i+} = m_{u,i-}$.

The mass matrix for the Down sector is obtained in a similar way.
%where now, in order to be sufficiently general, various parameters
%will carry the subscript $d$. 
The main difference between the two sectors is the ``universal'' 
strength which appears in fron of the matrix:
$g_{Y,u} \frac{v_u}{\sqrt{2}}$ for the Up sector and
$g_{Y,d} \frac{v_d}{\sqrt{2}}$ for the Down sector. The other
difference in the case of a non-hermitian matrix (Scenario (b))
is the splitting between ``left'' and ``right'' for each
family, which does not have to be the same for the two sectors.

%We have
%\begin{equation}
%\label{down}
%{\cal M}_{d} = g_{Y,d} \frac{v_{d}}{\sqrt{2}} \exp(i\theta_{d,ij}) \, ,
%\end{equation}
%where
%\begin{subequations}
%\label{angled}
%\begin{equation}
%\theta_{d,ii} = 0 \, ,
%\end{equation}
%\begin{equation}
%\theta_{d,ij} = \frac{(\Delta m_{d,i+,j-})^2}{4\,\Delta \mu_{d,ij}^2} \,.
%\end{equation}
%\end{subequations}
%The same scenario concerning the thickness of various domain walls
%as compared with the compactified dimension is applied here.

Notice that, in order to be more general, we aloow
the possibility of two different
mass scales: $v_{u}$ and $v_{d}$. If there were only {\em one} SM
Higgs field then $v_{u} = v_{d} =v$. In this case, the disparity
between the mass scales of the Up and Down sectors would come from
from the difference between $g_{Y,u}$ and $g_{Y,d}$, which, in turns,
could come from the differences between wave function overlaps, along
the fifth dimension, of the two sectors (modulo differences in the 
fundamental Yukawa couplings). To keep our discussions as general
as possible, we also allow for the possibility that two SM Higgs fields
exist.

It is beyond the scope of this paper to discuss in detail the
phenomenology of our model. It will be carried out elsewhere.

\subsubsection{Some Remarks on localization of
family domain walls along the sixth dimension}
\label{local}

In this section, we will briefly discuss one way to localize the
various domain walls responsible for ``fixing'' the phases of fermions
along the sixth dimension. There are probably several mechanisms
to achieve this. We will present one of such mechanisms, from the
point of view of effective field theory.

For simplicity, we shall assume in this section that $a_{i} =
a = f/\sqrt{\lambda/2}$. This simple assumption basically
refers to couplings between fermions and background scalar
fields which are invariant under the family symmetry.

First, let us list the parameters that we need to construct an 
almost pure phase mass matrix. From Section (\ref{HMM}),
we learned that we need: $\mu_{i}$ with $i=1,2,3$ which
control the thicknesses of the domain walls and $m_{i\pm}$ which
control the locations of the domain walls. We also learned
that, one can obtain a hermitian mass matrix when $m_{i+}=m_{i-}$
and a non-hermitian matrix when $m_{i+} \neq m_{i-}$.
It turns out to be a highly non-trivial task to find a mechanism
which can ``explain'' the origin of these parameters. In some
sense, it might even be overly ambitious to make such a claim.
We will, however, make an attempt to, at least, hint at one
possible scenario.

In Section (\ref{HMM}), we were basically doing the
``geography'' of family domain walls along the 
sixth dimension. To construct a
scenario for the ``geographical points'' (the various $m$'s), let us
recall that the family symmetry of our model is $S_{3}^{Q} \otimes
S_{3}^{U^c}$. The background scalar fields which couple to $Q$ or
$U^{c}$ will appear in terms such as $\bar{Q} \Phi Q$, $\bar{U^{c}}
\Phi U^{c}$. We will therefore need two of such background fields in
order to write down invariant Yukawa couplings: $\Phi_{Q}$ and
$\Phi_{U^{c}}$.  These background fields, $\Phi_{Q}$ and
$\Phi_{U^{c}}$, will be represented by $3 \times 3$ matrices. Some of
the details concerning the potential for these scalars are given in
Appendix \ref{appB}. Here, we will just quote the results. The
discussion below refers to the Up sector. As we have seen earlier, the
Down sector can be treated in exactly the same manner.

We will concentrate on scenario (a) of Section (\ref{HMM})
for the purpose of illustration.
We will assume the following Yukawa interactions:
\begin{equation}
\label{newyuk}
{\cal L}_Y = f \bar{Q} \Phi_{Q} Q + f \bar{U}^{c}  \Phi_{U^{c}}
U^{c} + h. c. \, ,
\end{equation}
where, for simplicity, we have put the two Yukawa couplings to
be equal. (A more general case can be accommodated straightforwardly.)
The minimization of the potential gives, at {\em tree level},
\begin{equation}
\label{QVEV}
\langle\,\Phi_{Q}\,\rangle =
\left(\begin{array}{ccc}
h_{1}(z)&0&0 \\
0&h_{2}(z)&0 \\
0&0&h_{3}(z)
\end{array}
\right)\ \, .
\end{equation}
One could assume that, at some deeper level and 
because of the family symmetry, the two background
fields behave in exactly the same manner, i.e. having similar
parameters, and , in consequence, one has
\begin{equation}
\label{UVEV}
\langle\,\Phi_{U^{c}}\,\rangle =
\left(\begin{array}{ccc}
h_{1}(z)&0&0 \\
0&h_{2}(z)&0 \\
0&0&h_{3}(z)
\end{array}
\right)\ \, .
\end{equation}
These VEV's will be shifted by radiative corrections. It is beyond
the scope of this paper to examine this problem and we will simply
parametrize these shifts by
\begin{equation}
\label{shift}
h_{i}(z) \rightarrow h_{i}(z) + \delta h_{i} \, ,
\end{equation}
where the shifts are assumed to be independent of $z$ and are
also assumed to be much smaller than $v_{i}$ (or $\mu_i$).

Combining Eq. (\ref{shift}) with Eq. (\ref{newyuk}), one can
make the following identification
\begin{equation}
\label{id0}
m_{i-} = m_{i+} = f \delta h_{i} \, .
\end{equation}
This is the case when one would obtain a hermitian mass matrix
of scenario (a) of Section (\ref{HMM})!
It goes without saying that there are two assumptions which
have been made. First, we have assumed the equality of the Yukawa
couplings in Eq. (\ref{newyuk}). Second, we have assumed that
the behavior of the two background scalar fields are identical.
These assumptions might come from some deeper symmetry between
$Q$ and $U^c$ (or $D^c$). This is very similar to the notion
of left-right symmetry that one encounters in four-dimensional
model building. In consequence, the hermiticity of the mass
matrix that we obtained by ``phenomenologically'' putting
$m_{i-} = m_{i+}$ might be justified by some form of left-right
symmetry.

In addition (\ref{shift}), one should also take into 
account vertex corrections
which will be different for $Q$ and $U^{c}$ (they have different
gauge interactions for example). Let us parametrize those shifts
by
\begin{subequations}
\begin{equation}
\label{Qshift}
\tilde{f}_Q = f + \delta f_{Q} \, ,
\end{equation}
\begin{equation}
\label{Ushift}
\tilde{f}_U = f + \delta f_{U^{c}} \, ,
\end{equation}
\end{subequations}
where the notations are self-explanatory. We will assume that
$\delta f_{Q,U^{c}} \ll f$. Naturally, $\tilde{f}_Q \neq
\tilde{f}_{U}$.

From the above equations, one can make the following identifications:
\begin{subequations}
\label{id}
\begin{equation}
m_{i-} = \tilde{f}_{Q} \delta h_{i} \, ,
\end{equation}
\begin{equation}
m_{i+} = \tilde{f}_{U} \delta h_{i} \, 
\end{equation}
\end{subequations}
Since one expects $\delta f_{Q} \neq \delta f_{U^{c}}$ and, in
consequence, $\tilde{f}_Q \neq \tilde{f}_U$, one would
expect, in general, $m_{i+} \neq m_{i-}$ which is a condition
for the appearance of a non-hermitian matrix. However,
an approximate hermitian matrix could arise if the radiative corrections
and, in particular, the difference in the radiative corrections
are small.
%In computing the elements of the mass matrix, one should keep in mind
%that the diagonal matrix elements are given by
%\begin{eqnarray}
%a_{ii} & = & \exp\left(i (\delta f_{Q} - \delta f_{U^{c}})\frac{\delta h_{i}^2}
%{\mu_{i}^2}\right) \, , \\ \nonumber
%       & \approx & 1 \, ,
%\end{eqnarray}
%where we have used the fact that $\left((\delta f_{Q} - \delta f_{U^{c}})
%(\delta h_{i}^2/\mu_{i}^2)\right) \ll 1$.
%From (\ref{id}), one can see that $m_{i-} \neq m_{i+}$, a condition
%for the existence of a non-hermitian matrix. 
One can see that, as
we turn off whatever interactions (gauge, etc...) which contribute
to the vertex corrections $\delta f_{Q,U^{c}}$, one recovers the
hermitian case, namely $m_{i-} = m_{i+}$.

Pursuing the same idea, one can also assume that $D^c$'s have
a similar coupling of the form $f \bar{D^{c}} \Phi_{D^{c}}
D^{c}$. Assuming that $\langle\,\Phi_{D^{c}}\,\rangle$ has a similar
form to Eqs. (\ref{QVEV},\ref{UVEV}), one can now see
that, in the absence of vertex corrections, one obtains
$m_{i-} = m_{i,u,+}=m_{i,d,+}$, which is just scenario
(a) discussed above. Since $D^c$ and $U^c$ have different
quantum numbers, one expects that their vertex corrections
will be different from each other. In consequence, one will 
obtain mass matrices of the form (\ref{PPMM1}) with coefficients
of the form (\ref{diag},\ref{offdiag}).
 
In the scenario just outlined above, one can make interesting
connections with the strong CP problem. In the absence of vertex
corrections, the mass matrix is {\em hermitian} and hence $\arg(\det
{\cal M}) =0$, a possible solution to the strong
CP~\cite{CP} problem?  (One could assume CP to be a
symmetry of the Lagrangian so that $\theta_{QCD} = 0$.) As mentioned
above, this hermiticity might come from some left-right symmetry ($Q
\leftrightarrow U^{c},D^{c}$) which gives $m_{i-} = m_{i+}$ at ``tree
level''. It could be quite provocative to see if there are
connections, if any, with previous solutions to the strong CP problem
which made use of the quintessential Left-Right
symmetry~\cite{Mohapatra:fy}.

Turning on the vertex corrections, the pure phase mass matrix
becomes non-hermitian and, as a consequence,
one would obtain a non-zero
contribution to the strong CP parameter $\bar{\theta}$. 
If this were truly a plausible scenario for the
strong CP problem, the resultant $\bar{\theta}$ should
obey the upper bound of $\sim 10^{-9}$. However,
it is
beyond the scope of this paper to analyze its magnitude. We will
come back to this issue in a subsequent paper. Our future
studies will focus on the following two questions. Will the
``radiative corrections'' be small enough so as to account for
both the phenomenological constraints on the mass matrices
{\em and} the magnitude of $\bar{\theta}$? If those
phenomenological constraints on the mass matrices require
a ``large'' radiative correction, is there a ``natural'' mechanism to
make $\bar{\theta}$ small enough?

%Two remarks are in order
%here. First, Reference \cite{branco} has carried out a detailed 
%analysis of pure phase mass matrices, showing how one can fit the
%observed pattern of quark masses and CKM matrix elements. Second,
%it was observed that one cannot, by a weak-basis transformation,
%bring a non-hermitian pure phase mass matrix to a hermitian one.
%This last remark is meant to emphasize that our two cases,
%hermitian and non-hermitian mass matrices, are distinct.

\section{Conclusion}

In this paper, we have studied the problem of fermion mass hierarchy
from the point of view of large extra dimensions.  To this end, we
have added two extra compact spatial dimensions.  In particular, we
have shown how one can construct a particular kind of mass matrices
which is very successful in fitting the pattern of quark masses and
mixing angles: The pure phase mass matrix. This matrix is
characterized by a universal Yukawa strength appearing in front of a
matrix whose elements are of the form $\exp (i \theta_{ij})$.  In our
construction, the universal Yukawa strength arises from the overlap of
the wave functions of the left-handed quarks (denoted by $Q$) and the
right-handed quarks (denoted by $U^c$ and $D^c$) along the fifth
spatial dimension ($y$).  Along $y$, all left-handed families are
localized at one place and all right-handed families at another place,
with the localization carried out by domain walls whose thicknesses
are assumed to be much smaller than the radius of compactification of
$y$. We then proceed to show that the almost pure phase mass matrix
arise from the overlap of wave functions between different families
and also between left-handed and right-handed quarks, along the sixth
dimension $z$. Along $z$, the ``phase determination'' 
is carried out by
domain walls whose thicknesses are assumed to be of the size of the
radius of compactification of $z$.

The almost-pure phase mass matrices obtained in six dimensions have
some interesting properties, according to the ``locations'' of
the family domain walls, which fix the phases, 
along the sixth dimension. In one case 
(scenario (a)) which is dubbed
``tree level'' in this paper, the domain walls for
$Q$ and $U^c$ or $D^c$ are
``located'' at the same place along the sixth dimension
$z$, for each family. The mass matrices thus obtained are
purely {\em hermitian}. In addition, apart from a different
universal Yukawa strength, the matrices of the Up and Down
sectors are identical, giving rise to a situation in which
the CKM matrix is simply a unit matrix. We then considered
a scenario in which either the domain wall for 
$U^c$ or $D^c$, or both, is 
split from that for $Q$. As we have shown 
in Section (\ref{HMM}), this
would imply that the mass matrix of at least one of the
two sectors is non-hermitian, and the two matrices will
be different from each other, implying a non-trivial
CKM matrix.

One should also keep in mind the possibility that the
mass matrices of both sectors are hermitian but not identical.
In this type of scenario, one would get a non-trivial CKM matrix 
as well as a correct spectrum. This possibility is mentioned
at the end of Section (\ref{HMM}). In order to be able
to build a model for the ``locations'' of various quarks
along the extra spatial dimensions, a full phenomenological
analysis of various possibilities should be carried out
to serve as a guidance. 
%A second case
%(scenario (b)) where all $Q$'s are ``located'' at one place
%and all $U^c$'s as well as all $D^c$'s are ``located''
%at another place, a hermitian mass matrix also arises,
%although the resulting CKM can now be different from
%the unit matrix. We then show that, by splitting
%$Q$'s from each other as well as from $U^c$'s and
%$D^c$'s, one obtains non-hermitian mass matrices which
%are most desirable from a phenomenological point of view.

These two cases of hermitian and non-hermitian mass
matrices might have important connections to the strong 
CP problem as we have briefly discussed above. This
interesting issue will be further investigated in a future
paper.

Finally, a number of interesting issues such as the Kaluza-Klein
modes, the extension to the lepton sector, and others will be
dealt with in future publications.

\begin{acknowledgments}
This work is supported in parts by the US Department
of Energy under grant No. DE-A505-89ER40518. One of us (PQH)
would like to thank the hospitalities of the theory groups
at the University of Rome, La Sapienza, and at Brookhaven
National Laboratory, where parts of this work were carried out.
We would like to thank Wai-Yee Keung, Paul Frampton, 
Mariano Quiros, David Dooling, and Andrea Soddu for discussions.
\end{acknowledgments}

\appendix
\section{}
\label{appA}
In this appendix we are going to present a brief review of the
spinorial representations of the orthogonal group, $O(D)$, in higher dimensions
($D>4$). We are going to follow closely the treatment done by Weinberg
in his book \cite{weinberg}, with a slightly different notation.

The starting point is a set of matrices, which spawn the Clifford
algebra, with the anticommutation relations
$\{\gamma_\mu,\gamma_\nu\}=2\delta_{\mu\nu}$. In addition, our
attention will be fixed on spaces of even dimensionality ($D=2n$) and,
subsequently, we'll extend it to odd dimension spaces.

Using the anticommutation relations of the $\gamma$ matrices, we can
define $n$ fermionic harmonic oscillators as
$a_i^+=\frac12\left(-\gamma_{2i}+i\gamma_{2i+1}\right)$ with
$i=0,\ldots,n-1$ that are independent and, therefore, the set of basis vectors of the representation space has $2^n$ elements which can be written as:
\begin{equation}
\label{state}
|s_1 s_2 \ldots s_n\rangle={a_1^+}^{s_1}{a_2^+}^{s_2}\cdots {a_n^+}^{s_n}
|0\rangle
\end{equation}
being $|0\rangle$ a vacuum annihilated by all destruction operators
$a_i$. In this basis the matrices $a_i$ take the form:
\begin{equation}
a_i=\begin{pmatrix}-1&0\\0&1\end{pmatrix}\otimes\cdots\otimes
\begin{pmatrix}-1&0\\0&1\end{pmatrix}\otimes\begin{pmatrix}0&1\\0&0\end{pmatrix}
\otimes I\!\!I_{2\times2}\otimes\cdots\otimes I\!\!I_{2\times2}
\end{equation}
the $-1$'s are due to the fact that $a_i^+$ and $a_j^+$ anticommute. Finally the
$\gamma$ matrices can be easily obtained and they read as:
\begin{equation}
\displaystyle
\setlength{\arraycolsep}{2pt}
\begin{array}{rcl}
\gamma_{2i}&=&-\sigma_3\otimes\cdots\otimes\sigma_3\otimes\sigma_1\otimes
I\!\!I_{2\times2}\otimes\cdots\otimes I\!\!I_{2\times2}\\
\gamma_{2i+1}&=&\sigma_3\otimes\cdots\otimes\sigma_3\otimes\sigma_2\otimes
I\!\!I_{2\times2}\otimes\cdots\otimes I\!\!I_{2\times2}
\end{array}
\end{equation}
where $\sigma_i$'s are the Pauli matrices. Note that this representation does not
give the usual representation in four dimensions but we can relate
both using the following unitary transformation:
\begin{equation}
U=\frac1{\left(\sqrt{2}\right)^n}(\sigma_3+\sigma_2)\otimes\cdots\otimes(\sigma_3+\sigma_2)
\end{equation}
so the ``usual'' representation is
\begin{equation}
\displaystyle
\setlength{\arraycolsep}{2pt}
\begin{array}{rcl}
\gamma_{2i}&=&\sigma_2\otimes\cdots\otimes\sigma_2\otimes\sigma_1\otimes
I\!\!I_{2\times2}\otimes\cdots\otimes I\!\!I_{2\times2}\\
\gamma_{2i+1}&=&\sigma_2\otimes\cdots\otimes\sigma_2\otimes\sigma_3\otimes
I\!\!I_{2\times2}\otimes\cdots\otimes I\!\!I_{2\times2}
\end{array}
\end{equation}

Since,
\begin{equation}
\gamma_{2i}\gamma_{2i+1}=i\,I\!\!I_{2\times2}\otimes\cdots\otimes 
I\!\!I_{2\times2}\otimes\sigma_2\otimes
I\!\!I_{2\times2}\otimes\cdots\otimes I\!\!I_{2\times2}
\end{equation}
the product of the $2n$ $\gamma$ matrices is:
\begin{equation}
\label{phaseres}
\prod_{i=0}^{2n-1}\gamma_i=i^n\sigma_2\otimes\cdots\otimes\sigma_2=\eta\,
\gamma_{2n}.
\end{equation}
Where $\eta$ is a phase such that $\gamma_{2n}\gamma_{2n}$ is the
identity.  Note that we are labeling the gamma matrix equivalent to
$\gamma_5$ with $2n$ instead $2n+1$. This difference comes from our
choice for the labeling starting from $0$ instead of $1$. This 
new matrix anticommutes with all
$\gamma$'s and therefore it implies that all spinorial representations
of $O(2n)$ are reducible.

Let us find now the spinorial representations of the orthogonal groups
with odd dimensionality, $D=2n+1$; this is much more simpler once we
have the representation for $O(2n)$ since we just have to take this
representation and add the $\gamma_{2n}$ matrix. In this case the
representation is irreducible because we can not find any independent matrix that
anticommutes with all the gamma matrices.

The transition of the $O(D)$ representations to $O(D-1,1)$ representations
is done through a wick rotation.

To finalize we will explicitly write the gamma matrices for $O(5,1)$.

\begin{itemize}
\item{Six dimensions (with metric $(-+++++)$):
\begin{equation}
\setlength{\arraycolsep}{2pt}
\begin{array}{c}
\begin{array}{rclcrcl}
\Gamma_0&=&i\,\sigma_2\otimes\sigma_1\otimes I\!\!I_{2\times2}=
\begin{pmatrix}
0_{4\times4}&-\gamma_0\\
\gamma_0&0_{4\times4}
\end{pmatrix}
&\qquad\qquad&
\Gamma_1&=&\sigma_2\otimes\sigma_2\otimes\sigma_1=
\begin{pmatrix}
0_{4\times4}&-\gamma_1\\
\gamma_1&0_{4\times4}
\end{pmatrix}\\[.5truecm]
\Gamma_2&=&\sigma_2\otimes\sigma_2\otimes\sigma_2=
\begin{pmatrix}
0_{4\times4}&-\gamma_2\\
\gamma_2&0_{4\times4}
\end{pmatrix}&&
\Gamma_3&=&\sigma_2\otimes\sigma_2\otimes\sigma_3=
\begin{pmatrix}
0_{4\times4}&-\gamma_3\\
\gamma_3&0_{4\times4}
\end{pmatrix}\\[.5truecm]
\Gamma_y&=&\sigma_2\otimes\sigma_3\otimes I\!\!I_{2\times2}=
\begin{pmatrix}
0_{4\times4}&-i\,\gamma_5\\
i\,\gamma_5&0_{4\times4}
\end{pmatrix}&&
\Gamma_z&=&\sigma_1\otimes I\!\!I_{2\times2}\otimes I\!\!I_{2\times2}=
\begin{pmatrix}
0_{4\times4}&I\!\!I_{4\times4}\\
I\!\!I_{4\times4}&0_{4\times4}
\end{pmatrix}
\end{array}\\[2truecm]
\Gamma_7=\sigma_3\otimes I\!\!I_{2\times2}\otimes 
I\!\!I_{2\times2}=
\begin{pmatrix}
I\!\!I_{4\times4}&0_{4\times4}\\
0_{4\times4}&-I\!\!I_{4\times4}
\end{pmatrix}
\end{array}
\end{equation}
}
\end{itemize}

Notice that, in the above equations, $\gamma^{\mu}$ ($\mu =0,1,2,3$) 
and $\gamma_5$ are simply defined {\em here} as
$\gamma_0 = \sigma_1\otimes I\!\!I_{2\times2}$,
$\gamma_{i} = i\,\sigma_2\otimes\sigma_{i}$,
$\gamma_{5} = \sigma_3\otimes I\!\!I_{2\times2}$.
These definitions just happen to coincide with 
the 4-dimensional ones with
a metric $(+---)$. This is simply a compact way of writing
the 6-dimensional $\Gamma$'s. There is {\em no change} in metric.
To see how the 6-dimensional metric $(-+++++)$ reduces to a
4-dimensional metric $(-+++)$ when the two extra spatial dimensions
are compactified, one rewrites $\gamma^{\mu}$
in terms of the gamma matrices which correspond to the metric
$(-+++)$, namely $\tilde{\gamma}^{\mu} = i\, \gamma^{\mu}$.
In this way, the kinetic terms will be preceded with a plus sign
when they are reexpressed in terms of $\tilde{\gamma}^{\mu}$. Of course,
$\gamma_5$ remains unchanged.

\section{}
\label{appB}
In this article, we had been discussing models in
which fermions are localized at one place or another along the
extra dimensions and inside fat branes with the same or different
widths and how these settings could affect the phenomenology of the 4D
models. However, we did not provide any model that explains these
different settings; this will be addressed in this appendix.

As an example we are going to study the possibility that the background
scalar field is 
a composite of fields that transforms under a three dimensional representation
of the family group; therefore this background scalar
 field $\Phi$ takes the form:
\begin{equation}
\label{compos}
\Phi(x)=(\phi_1(x)\,\phi_2(x)\,\phi_3(x))^+\otimes
\begin{pmatrix}\phi_1(x)\\ \phi_2(x)\\ \phi_3(x)
\end{pmatrix}
\end{equation}
where the $\phi_i$'s are the ``fundamental fields'' from which the background
scalar field is composed.

The first of the models we are going to propose consists on a $\phi^4$
potential, without cubic terms, for $2$ composite fields of the form
(\ref{compos}),
\begin{equation}
\setlength{\arraycolsep}{2pt}
\displaystyle
\begin{array}{rcl}
V(\Phi_1,\Phi_2)&=&\displaystyle\frac{m_1^2}2\text{Tr}[\Phi_1\Phi_1^+]+
\frac{m_2^2}2\text{Tr}[\Phi_2\Phi_2^+]+\frac{m_3}2(\text{Tr}[\Phi_1\Phi_2^+]
+\text{h.c.})\\[.25truecm]
&+&\displaystyle\frac{\lambda_1}4\,\text{Tr}[(\Phi_1\Phi_1^+)^2]+
\frac{\lambda_2}4\,\text{Tr}[(\Phi_2\Phi_2^+)^2]+\frac{\lambda_3}2\,
\text{Tr}[(\Phi_1\Phi_1^+)(\Phi_2\Phi_2^+)]\\[.25cm]
&+&\displaystyle\frac{\lambda_4}4\left(
\text{Tr}[\Phi_1\Phi_2^+\Phi_1\Phi_2^+]+\text{h.c.}\right)\\[.25truecm]
&=&\displaystyle\frac{m_1^2}2|\Phi_1|^2+\frac{m_2^2}2|\Phi_2|^2+
m_3|\Phi_1||\Phi_2|\cos^2\alpha\\[.25truecm]
&+&\displaystyle\frac{\lambda_1}4\,|\Phi_1|^2+
\frac{\lambda_2}4\,|\Phi_2|^2+\frac{\lambda_3}2\,|\Phi_1|^2
|\Phi_2|^2\cos^2\alpha\\[.25cm]
&+&\displaystyle\frac{\lambda_4}2\,|\Phi_1|^2|\Phi_2|^2\cos^4\alpha
\end{array}
\end{equation}
where $m_{1,2}^2$ are negative coefficients. We also made use of the following
definitions,
\setlength{\arraycolsep}{2pt}
\begin{eqnarray}
|\Phi|^2&=&\text{Tr}[\Phi\Phi^+]\\[.25truecm]
\cos^2\alpha&=&\,\text{Re}\left(\frac{\text{Tr}[\Phi_1\Phi_2^+]}
{|\Phi_1||\Phi_2|}\right)
\end{eqnarray}

If $2\,m_3>-(\lambda_3+\lambda_4\cos^2\alpha)|\Phi_1||\Phi_2|$ holds
then the minimum of the potential occurs when both fields take
expectation values along orthogonal directions, that is, when 
$\text{Tr}[\Phi_1\Phi_2^+]=0$. In consequence, if we suppose that
the expectation values for $|\Phi_1|$ and $|\Phi_2|$ are $u$ and $v$
respectively, the potential we have to minimize is:
\begin{equation}
V(u,v)=\frac{m_1^2}2u^2+\frac{m_2^2}2v^2+\frac{\lambda_1}4u^4+
\frac{\lambda_2}4v^4
\end{equation}
and its minimum is located at:
\begin{equation}
\setlength{\arraycolsep}{2pt}
\begin{array}{rclcrcl}
u&=&\displaystyle\sqrt{-\frac{m_1^2}{\lambda_1}}&\qquad
&v&=&\displaystyle\sqrt{-\frac{m_2^2}{\lambda_2}}
\end{array}
\end{equation}
therefore we can suppose that,
\setlength{\arraycolsep}{2pt}
\begin{eqnarray}
\langle\,\Phi_1\,\rangle&=&
\setlength{\arraycolsep}{8pt}
\begin{pmatrix}
u&0&0\\
0&0&0\\
0&0&0
\end{pmatrix}\\
\langle\,\Phi_2\,\rangle&=&
\setlength{\arraycolsep}{8pt}
\begin{pmatrix}
0&0&0\\
0&v&0\\
0&0&0
\end{pmatrix}.
\end{eqnarray}
This means that this model will localize the components of the family
multiplet at different positions along the sixth dimension, namely
$0$, $u$ and $v$.

We can extend this model to localize all the components of the family 
multiplet inside the orbifold, with the background scalar
field in (\ref{yuktilde}) having
all three eigenvalues different from zero. This can be done 
if we add another background field that can 
be a singlet, or a composite like the ones used. In the former case the three 
components will be shifted by the same amount, $s$, ending in positions
$s+u$, $s+v$ and $s$; in the later case the fermions will be located at 
$u$, $v$ and $z$ (the expectation value for the third field).

% now the references. delete or change fake bibitem. delete next three
%   lines and directly read in your .bbl file if you use bibtex.

% \begin{figure}
% \caption{}
% \label{}
% \end{figure}

% \begin{table}
% \caption{}
% \label{}
% \end{tabular}
% \end{table}


\begin{thebibliography}{99}
\bibitem{branco}{G.~C.~Branco, J.~I.~Silva-Marcos and M.~N.~Rebelo,
Phys.\ Lett.\ B {\bf 237} 446 (1990); G.~C.~Branco and J.~I.~Silva-Marcos,
Phys.\ Lett.\ B {\bf 359}, 166 (1995); G.~C.~Branco, D.~Emmanuel-Costa and 
J.~I.~Silva-Marcos, Phys.\ Rev.\ D {\bf 56}, 107 (1997).}
%%CITATION = PHLTA,B237,446;%%
%%CITATION = HEP-PH 9507299;%%
%%CITATION = HEP-PH 9608477;%%

%\cite{Fishbane:xa}
\bibitem{Fishbane:xa}
P.~M.~Fishbane and P.~Q.~Hung,
%``Dynamics And Symmetries For Pure Phase Mass Matrices,''
Phys.\ Rev.\ D {\bf 57}, 2743 (1998).
%%CITATION = PHRVA,D57,2743;%%

\bibitem{dmm}{See e.g.
P.~Kaus and S.~Meshkov, Phys.\ Rev.\ D {\bf 42} 1863 (1990), and
references therein.}
%%CITATION = PHRVA,D42,1863;%%

%\cite{Arkani-Hamed:1998rs}
\bibitem{Arkani-Hamed:1998rs}
N.~Arkani-Hamed, S.~Dimopoulos and G.~R.~Dvali,
%``The hierarchy problem and new dimensions at a millimeter,''
Phys.\ Lett.\ B {\bf 429}, 263 (1998)
[arXiv:hep-ph/9803315].
%%CITATION = HEP-PH 9803315;%%

%\cite{Antoniadis:1990ew}
\bibitem{Antoniadis:1990ew}
I.~Antoniadis,
%``A Possible New Dimension At A Few Tev,''
Phys.\ Lett.\ B {\bf 246}, 377 (1990).
%%CITATION = PHLTA,B246,377;%%

%\cite{Arkani-Hamed:1999dc}
\bibitem{Arkani-Hamed:1999dc}
N.~Arkani-Hamed and M.~Schmaltz,
%``Hierarchies without symmetries from extra dimensions,''
Phys.\ Rev.\ D {\bf 61}, 033005 (2000)
[arXiv:hep-ph/9903417].
%%CITATION = HEP-PH 9903417;%%

%\cite{Hoyle:2000cv}
\bibitem{Hoyle:2000cv}
C.~D.~Hoyle, U.~Schmidt, B.~R.~Heckel, E.~G.~Adelberger, J.~H.~Gundlach, D.~J.~Kapner and H.~E.~Swanson,
%``Sub-millimeter tests of the gravitational inverse-square law: A search  for 'large' extra dimensions,''
Phys.\ Rev.\ Lett.\  {\bf 86}, 1418 (2001)
[arXiv:hep-ph/0011014].
%%CITATION = HEP-PH 0011014;%%

\bibitem{CP}{R.~D.~Peccei and H.~R.~Quinn, Phys.\ Rev.\ D 
{\bf 16}, 1791 (1977); H.~Y.~Cheng, Phys.\ Rept.\  {\bf 158}, 1 (1988);
J.~E.~Kim, Phys.\ Rept.\  {\bf 150}, 1 (1987). For recent reviews see, for 
example, R.~D.~Peccei, hep-ph/9807514; M.~Dine, hep-ph/0011376; 
H.~R.~Quinn, hep-ph/0110050, and references therein.}
%%CITATION = PHRVA,D16,1791;%%
%%CITATION = PRPLC,158,1;%%
%%CITATION = PRPLC,150,1;%%
%%CITATION = HEP-PH 9807514;%%
%%CITATION = HEP-PH 0011376;%%
%%CITATION = HEP-PH 0110050;%%

%\cite{Teshima:1996xc}
\bibitem{Teshima:1996xc}
T.~Teshima and T.~Sakai,
%``Violation of universal Yukawa coupling and quark masses,''
Prog.\ Theor.\ Phys.\  {\bf 97}, 653 (1997)
[arXiv:hep-ph/9608447].
%%CITATION = HEP-PH 9608447;%%

%\cite{Georgi:2000wb}
\bibitem{Georgi:2000wb}
H.~Georgi, A.~K.~Grant and G.~Hailu,
%``Chiral fermions, orbifolds, scalars and fat branes,''
Phys.\ Rev.\ D {\bf 63}, 064027 (2001)
[arXiv:hep-ph/0007350].
%%CITATION = HEP-PH 0007350;%%

%\cite{Cheng:1999bg}
\bibitem{Cheng:1999bg}
H.~C.~Cheng, B.~A.~Dobrescu and C.~T.~Hill,
%``Electroweak symmetry breaking and extra dimensions,''
Nucl.\ Phys.\ B {\bf 589}, 249 (2000)
[arXiv:hep-ph/9912343].
%%CITATION = HEP-PH 9912343;%%

%\cite{Kaplan:1992bt}
\bibitem{Kaplan:1992bt}
D.~B.~Kaplan,
%``A Method for simulating chiral fermions on the lattice,''
Phys.\ Lett.\ B {\bf 288}, 342 (1992)
[arXiv:hep-lat/9206013].
%%CITATION = HEP-LAT 9206013;%%

%\cite{Mohapatra:fy}
\bibitem{Mohapatra:fy}
R.~N.~Mohapatra and G.~Senjanovic,
%``Natural Suppression Of Strong P And T Noninvariance,''
Phys.\ Lett.\ B {\bf 79}, 283 (1978).
%%CITATION = PHLTA,B79,283;%%

\bibitem{weinberg}{S.~Weinberg, \textit{The Quantum Theory of Fields III},
Cambridge U.P. (2000).}
\end{thebibliography}
\end{document}